\documentclass[english,aps,preprint,nofootinbib,
superscriptaddress,prd,preprintnumbers,
]{revtex4-1}
\pdfoutput=1
\usepackage{multirow}
\usepackage[usenames,dvipsnames,svgnames,table]{xcolor}
\usepackage{mathrsfs}
\usepackage{verbatim}
\usepackage{tikz}
\usepackage{pgfplots}
\pgfplotsset{compat=newest,every axis plot/.append style={line width=1pt}}
\usepackage{babel}
\usepackage{float}
\usepackage{graphicx}
\usepackage{svg}
\usepackage{hyperref}
\usepackage{amsthm,amsmath,amssymb,amsfonts}
\usepackage{url}
\usepackage{xspace}
\usepackage{lipsum}
\usepackage{enumitem}

\definecolor{lightgray}{gray}{0.9}
\definecolor{Amber}{rgb}{1.0, 0.75, 0.0}
\definecolor{blizzardblue}{rgb}{0.67, 0.9, 0.93}
\definecolor{myblue}{HTML}{0273B2}
\colorlet{myblueLight}{myblue!35}

\usepackage{tikz}

\usepackage{epsfig}
\newcommand{\postscript}[2]{\setlength{\epsfxsize}{#2\hsize}
   \centerline{\epsfbox{#1}}}

\definecolor{orange}{cmyk}{0,0.5,1,0}
\definecolor{rossoCP3}{cmyk}{0,.88,.77,.40}
\definecolor{graa}{rgb}{0.8,0.8,0.8}
\definecolor{blaa}{rgb}{0.2,0.2,0.6}

\begin{document}
\count\footins = 1000

\preprint{MPP-2026-84}

\vskip1cm

\title{\color{rossoCP3} 
Breaking Free from the Swampland of Impossible Universes through the DESI Portal}

\author{\bf Luis A. Anchordoqui}

\affiliation{Department of Physics and Astronomy,  Lehman College, City University of
  New York, NY 10468, USA
}

\affiliation{Department of Physics,
 Graduate Center,  City University of
  New York,  NY 10016, USA
}

\affiliation{Department of Astrophysics,
 American Museum of Natural History, NY
 10024, USA
}

\author{\bf Dieter\nolinebreak~L\"ust}

\affiliation{Max--Planck--Institut f\"ur Physik,  
 Werner--Heisenberg--Institut, 85748 Garching, Germany}

\affiliation{Arnold Sommerfeld Center for Theoretical Physics, 
Ludwig-Maximilians-Universit\"at M\"unchen,
80333 M\"unchen, Germany
}

\begin{abstract}
  \noindent The persistent challenge of creating stable de Sitter vacua within string theory undermines the observational validity of the  $\Lambda$ cold dark matter (CDM) model. This difficulty suggests that the  concordance model of cosmology, characterized by a constant dark energy  $\Lambda$, may reside in the {\it swampland} of inconsistent quantum gravity theories rather than the string {\it landscape} of consistent ones.  Recent observational data, particularly from the Dark Energy Spectroscopic Instrument (DESI), have significantly challenged $\Lambda$CDM cosmology. Specifically, the combination of DESI baryon acoustic oscillation measurements with cosmological surveys seem to indicate a preference for a dynamic, time-evolving dark energy rather than a constant, with roughly 10\% reduction in density over the last several billion years. This review summarizes significant advancements made over the past two years in linking DESI findings to string-inspired scenarios.
  \end{abstract}

\maketitle

\tableofcontents

\newpage

\section{Introduction}

The nature of the present day acceleration of the cosmic expansion~\cite{SupernovaSearchTeam:1998fmf,SupernovaCosmologyProject:1998vns},
generally known as dark energy, has proved exceedingly challenging to
understand theoretically. On the one hand, the simplest scenario in which dark energy is a cosmological constant $\Lambda$ can accommodate nearly all experimental data, with  measurements of the cosmic microwave background
(CMB) providing strong evidence for a description of the universe on large scales and its
evolution history in terms of the $\Lambda$ cold dark matter (CDM)
model~\cite{Efstathiou:1990xe,WMAP:2012nax,Planck:2018vyg}.  On the other hand, in $\Lambda$CDM the positive vacuum energy per unit volume must be meticulously
fine-tuned to an extremely small value ($\Lambda \sim 10^{-120}$ in reduced Planck units) in the early universe relative to the energy
scale at that time such that matter domination and structure formation
could occur. On that account, $\Lambda$CDM cosmology does not provide an
explanation but at best an effective description of the dark
energy.

Deepening the dark energy mystery, the latest findings on baryon
acoustic oscillations (BAO) from the Dark Energy
Spectroscopic Instrument (DESI) data release 2 (DR2) combined with CMB information and Type Ia supernova (SN Ia) datasets suggest a preference for dynamical dark
energy over the cosmological constant~\cite{DESI:2025zgx}. Specifically, DESI DR2 combined with these cosmological surveys hint that dark energy's density may have decreased by roughly 10\% over the last several billion years.

Adding to the story, the case for dynamical dark energy remains robust while considering the combination SN Ia and BAO data with the cross-correlation between DESI luminous red galaxies and CMB lensing~\cite{Sabogal:2025jbo}. Besides, integrating BAO data from the Dark Energy Survey (DES)~\cite{DES:2024pwq} to the joint analysis further strengthens the evidence against $\Lambda$CDM~\cite{Ishak:2025cay}. All in all, while not yet statistically conclusive, it is intriguing to explore the possibility that DESI data correspond to a
real signal of physics beyond $\Lambda$CDM. One simple, but theoretically well motivated mechanism for dynamical
dark energy is a noninteracting scalar field, generally called quintessence, rolling down its
potential~\cite{Peebles:1987ek,Ratra:1987rm,Wetterich:1994bg,Caldwell:1997ii}. As might be expected, the phenomenology of quintessence dark energy depends upon the choice of potential.

The last chapter of this story is not courtesy of experiment but instead powered by theory. The Swampland program seeks to identify universal principles that distinguish which low-energy effective
field theories (EFTs) are consistent with nonperturbative quantum gravity
considerations~\cite{Vafa:2005ui}. Essentially, it explores the boundaries between the string 
landscape (i.e. the area populated by EFTs compatible with quantum
gravity) and the swampland (i.e. the area inhabited by
consistent-looking low-energy EFTs that are incompatible with quantum
gravity). It is self-evident that the swampland is wider than the
landscape, and actually it
surrounds the landscape. The guiding principles for fencing off
the landscape have been dubbed swampland conjectures. There are many
conjectures in the market, indeed too many to all be listed here; for reviews see e.g.,~\cite{Palti:2019pca,vanBeest:2021lhn,Grana:2021zvf,Agmon:2022thq,Anchordoqui:2024ajk}. Obviously, to ensure consistency with a UV-complete theory of quantum gravity quintessence field potentials must satisfy constraints imposed by swampland conjectures. This review summarizes major progress made over the past two years in bridging DESI observational data with swampland constraints.\footnote{This review focuses on quintessence models driven by findings from DESI data; for broader overviews of dark energy within string theory, see e.g.~\cite{Copeland:2006wr,Andriot:2026lac}.}

The layout is as follows. We begin in Sec.~\ref{sec:2} with a general survey of the available data, outlining
the motivations and caveats associated with dynamical dark energy. In Sec.~\ref{sec:3} we review pertinent swampland conjectures and analyze the mechanisms by which the low-energy effective theory inherits properties from the overarching string theory. In Sec.~\ref{sec:4} we examine the various string-inspired quintessence models that have been proposed to accommodate DESI findings and we confront them with generic cosmological requirements from swampland conjectures. In Sec.~\ref{sec:5}, we analyze interacting quintessence models that feature a non-gravitational coupling to dark matter. Lastly, in Sec.~\ref{sec:6}
we provide our final thoughts.

\section{Mapping Dark Energy with DESI}
\label{sec:2}

There are two basic hypotheses in modern cosmology: {\it (i)}~the
assumption of the validity of general
relativity~\cite{Einstein:1916vd} and {\it (ii)}~the cosmological
principle, which is the assumption that the universe on cosmological
scales is homogeneous and isotropic~\cite{Einstein:1917ce}.\footnote{See, however,~\cite{Secrest:2025wyu}.} Together, these two hypotheses set constraints on the four-dimensional (4D) spacetime metric, which reduces to the maximally-symmetric Friedmann-Lama\^{\i}tre-Robertson-Walker (FLRW) line element
\begin{equation}
  ds^2 = -dt^2 + a^2(t) \left[ \frac{dr^2}{1-kr^2} + r^2 (d\vartheta^2 + \sin^2 \varphi \ d\varphi^2) \right] \,,
\label{metric}
\end{equation}
where $(t, r, \vartheta, \varphi)$ are comoving coordinates,  $k (= -1, 0, 1)$ parametrizes the curvature of the homogeneous and isotropic spatial sections, and $a(t)$ is the cosmic scale factor, which is related to the redshift $z$ by $a = 1/(1+z)$~\cite{Friedmann:1922kd,Friedmann:1924bb,Lemaitre:1927zz,Robertson:1935zz,Walker:1937qxv}. The expansion rate of the Universe is measured by the Hubble parameter,
\begin{equation}
  H = \frac{\dot {a}}{a} \, ,
\label{HubbleP}
\end{equation}  
and its present-day value is known as the Hubble constant $H_0 = 100~h~{\rm km/Mpc/s}$, with $0< h < 1$~\cite{Hubble:1929ig}.

Modern cosmological observations  strongly support a spatially flat $(k=0)$ universe composed of approximately 68-70\% dark energy and 30-32\% matter (the vast majority of which is dark matter)~\cite{ParticleDataGroup:2024cfk}. Throughout this review we assume $(k=0)$, unless otherwise specified. Dark energy can be characterized as a perfect fluid with energy density $\rho_{de}$, pressure $P_{de}$, and equation of state
\begin{equation}
  w \equiv P_{de}/\rho_{de} < - 1/3 \, .
\end{equation}  
Note that when substituting $w < -1/3$ into the second Friedmann equation
\begin{equation}
  \frac{\ddot a}{a} = - \frac{1}{6 M_p^2} \ \biggl(\rho_{m } + 2 \rho_{r} + \rho_{de} (1 + 3 w) \biggr)
\label{Frie2}
\end{equation}    
it causes cosmic acceleration, where $\rho_m$ is the nonrelativistic
matter density, $\rho_r$ the radiation density, and $M_p$ the reduced
Planck mass. For convenience, the present-day densities $\rho_{de,
  {\rm today}}$, $\rho_{m,{\rm today}}$, $\rho_{r,{\rm today}}$,
$\rho_{b,{\rm today}}$ and $\rho_{k,{\rm today}}$ are sometimes expressed as density parameter ratios 
$\Omega_i = \rho_{i,{\rm today}}/\rho_{\rm crit}$ of the $i$ component
$\rho_{i,{\rm today}}$ to the critical density required for a flat
universe $\rho_{\rm crit} = 3 H_0^2 M_p^2$, with $\rho_{b,{\rm
    today}}$ the baryon density and $\rho_{k,{\rm
    today}} = 3 M_P^2 k/a_{\rm today}$  the effective curvature
density.

For a cosmological constant $\Lambda$, which represents a constant vacuum energy density, $w=-1$. The Chevallier-Polarski-Linder (CPL) parametrization is a two-parameter, first-order Taylor expansion method, 
\begin{equation}
  w(z) = w_0 + w_a \ \frac{z}{(1 + z)} \,,
\end{equation}
defining the dark energy equation of state as a function of redshift~\cite{Chevallier:2000qy,Linder:2002et}. The CPL parametrization is widely used as a benchmark for testing dynamical dark energy against the standard cosmological constant ($w_0 = -1,w_a =0$).

DESI is an experiment that was designed to map the equation of state of dark energy using BAOs as a cosmic standard ruler~\cite{Weinberg:2013agg}. In the early universe, the interaction between gravitational collapse and photon pressure created oscillations (sound waves) within the hot plasma of photons and baryons. At the epoch of recombination, photons decoupled from matter forming CMB, while the baryons remained frozen into spherical shells. This process imprinted a characteristic scale known as the sound crossing horizon
\begin{equation}
  r_d = \int_{z_d}^\infty \frac{c_s(z)}{H(z)} dz \, ,
\end{equation}
where $c_s(z)$ is the speed of sound in the photon-baryon fluid and $z_d \approx 1060$ is the redshift at which acoustic waves stall because photons no longer {\it drag} the baryons. Assuming standard pre-recombination physics, the sound horizon can be computed given the densities of baryons, CDM, photons, and other relativistic species, yielding $r_d \sim 147~{\rm Mpc}$~\cite{Brieden:2022heh}. By measuring this fixed physical scale in the distribution of galaxies, DESI can precisely track the expansion history of the universe. Indeed, DESI surveys positions, distances, and redshifts of galaxies and quasars (in the first year it collected about 6 million objects~\cite{DESI:2024aax,DESI:2024uvr}, by the third year it collected 30 milion~\cite{DESI:2025zgx}, and by the fifth year it is expected to measure 40 million).

BAO surveys measure two distance ratios:  
\begin{equation}
  \theta_{\rm BAO} = r_d/D_M(z) \, ,
\end{equation}  
and 
\begin{equation}
\Delta z_{\rm BAO} = r_d/D_H  \, ,
\end{equation}  
where
\begin{equation}
  D_M (z)  =  \left\{ \begin{array}{ll} \dfrac{1}{H_0 \sqrt{\Omega_k}} \sinh \left[\sqrt{\Omega_k}  \displaystyle \int_0^z \dfrac{H_0\ dz'}{H(z')} \right] & ~~~~~k = -1 \\
                        \displaystyle \int_0^z \dfrac{dz'}{H(z')} & ~~~~~k =0 \\
\dfrac{1}{H_0 \sqrt{|\Omega_k|}} \sin \left[\small{H_0 \ \sqrt{|\Omega_k|}}  \displaystyle \int_0^z \dfrac{dz'}{H(z')} \right] & ~~~~~k = +1 
\end{array}
\right.                                 \end{equation}
is the comoving angular diameter distance to a specific redshift $z$ and 
\begin{equation}
  D_H (z) = \frac{1}{H(z)}
\end{equation}  
is the Hubble distance~\cite{Hogg:1999ad}.\footnote{The comoving angular diameter distance is the transverse comoving distance that relates the physical size ($S_{\rm phys}$) of an object to its observed angular size ($\Delta \theta$) on the sky via the relation $\Delta \theta = S_{\rm com}/D_M$,  where $S_{\rm com} = S_{\rm phys} (1+z)$
 is the comoving size.} DESI maps 3D structure by isolating radial (line-of-sight) and angular (transverse) modes. On the one hand, by characterizing the angular separation of galaxy clustering ($\theta_{\rm BAO}$) across the sky, and combining it with the known size of 
 of the sound crossing horizon, $D_M(z)$ can be determined. On the other hand, a measurement of the characteristic separation scale in redshift space along the line of sight allows for the direct determination of 
of the Hubble parameter
 \begin{equation}
H(z) = H_0 \sqrt{\Omega_m (1+z)^3 + \Omega_r (1+z)^4 + \Omega_k (1+z)^2 + \Omega_{de} \, {\cal E} (z, w)} \,, 
\end{equation}
where 
\begin{equation}
  {\cal E} = \frac{\rho_{de}(z)}{\rho_{de,{\rm today}}} = \exp \left[ 3 \int_0^z (1 + w(z')) \frac{dz'}{1+z'} \right] \, .
\end{equation}
$D_M(z)$ and $H(z)$ encode the expansion history of the Universe.

The DESI Collaboration
analyzed the $(w_0,w_a)$ plane, assuming that the equation of state of
dark energy satisfies the CPL parametrization. For theoretical cosmology calculations, the collaboration adopted the Code for Anisotropies in the Microwave Background ({\tt CAMB}) Boltzmann solver~\cite{Lewis:1999bs}. The Markov Chain Monte Carlo (MCMC) {\tt COBAYA}
software~\cite{Torrado:2020dgo} was employed for analyses to infer cosmological parameters from the data. Spectra were analyzed with the {\tt REDROCK} algorithm to determine redshifts (see Sec. 4.1 in~\cite{eBOSS:2020mzp}). In this section, we present a brief overview of DESI findings on $w(z)$.

\subsection{DESI DR1}

\begin{table}
  \caption{Evolution of $\Lambda$CDM rejection significance over the
    years for mixed dataset compilations. The  results listed in the
    first column are based on DESI DR1, CMB, and three SN Ia datasets
    (PantheonPlus~\cite{Scolnic:2021amr,Brout:2022vxf},
    Union3~\cite{Rubin:2023jdq}, and
    DESY5~\cite{DES:2024jxu})~\cite{DESI:2024mwx}.  The  results
    listed in the second column are based on DESI DR2 and the same datasets for CMB and SN Ia~\cite{DESI:2025zgx}. The results listed in the third column are based on  DESI DR2, CMB data, and SN Ia recalibrations (DES-Dovekie~\cite{DES:2025sig}, Union3.1, and PantheonPlus with corrected host mass estimates for low-redshift)~\cite{Hoyt:2026fve,Rubin:2026qdt}. \label{tabla1}}
\renewcommand{\arraystretch}{1.3}
\rowcolors{1}{}{lightgray}
    \centering
    \setlength\tabcolsep{0pt}
    \begin{tabular}{ |c|c|c|c| }
    \hline
      \rowcolor{myblueLight}
      Dataset Compilations & DESI DR1 & DESI DR2 & ~~~SN-recalibration~~~ \\
\hline
      ~~~~~~DESI+CMB~~~~~~ & $2.6\sigma$ & $3.1\sigma$ & $3.1\sigma$ \\
      ~~~~~DESI+CMB+PantheonPlus~~~~~ & ~~~~~~~~~~$2.5\sigma$~~~~~~~~~~ & ~~~~~~~~~~$2.8\sigma$~~~~~~~~~~ & ~~~~~~~~~~$3.2\sigma$~~~~~~~~~~ \\
      ~~~DESI+CMB+Union~~~ & ~~~$3.5\sigma$~~~ & ~~~$3.8\sigma$~~~ & ~~~$3.3\sigma$~~~\\
      ~~~DESI+CMB+DES~~~ & ~~~$3.9\sigma$~~~ & ~~~$4.2\sigma$~~~ & ~~~$3.3\sigma$~~~\\
\hline
    \end{tabular}
 
  \end{table}

In 2024, the DESI Collaboration reported the analysis of DR1~\cite{DESI:2024mwx}. The MCMC 
likelihood analysis based on DESI BAO data alone lacks the precision to break the $w_0$-$w_a$ degeneracy, causing the results to be limited by prior assumptions,
\begin{equation}
  \left. \begin{array}{ccl}
           w_0 & = & -0.55^{+0.39}_{-0.21} \\
           w_a & < & -1.32 \\
           \end{array} \right\} {\rm DESI \ BAO} \,,
\end{equation}
with the upper bound on $w_a$ referring to the 68\% limit.

Subsequently, the DESI DR1 was integrated with CMB data, including measurements of the temperature (TT), polarization (EE), and cross (TE) power spectra by the Planck spacecraft~\cite{Planck:2019nip}, as well as combination of CMB lensing data from the Atacama Cosmology Telescope (ACT)~\cite{ACT:2023dou,ACT:2023ubw,ACT:2023kun} and Planck’s PR4 maps~\cite{Carron:2022eyg}.
The combined result of the DESI data set with CMB measurements gives
\begin{equation}
  \left. \begin{array}{ccl}
           w_0 & = & -0.45^{+0.34}_{-0.21} \\
           w_a & = & -1.79^{+0.48}_{-1.00} \\
           \end{array} \right\} {\rm DESI \ BAO + CMB} \,,
\end{equation}
indicating a preference for an evolving dark energy equation of state at the  $2.6\sigma$ level. 

Finally, three distinct SN Ia datasets were utilized to further break the degeneracy in the $(w_0, w_a)$ plane.
The PantheonPlus compilation, which  consists of 1550 spectroscopically-confirmed SN Ia in the redshift range $0.001 < z < 2.26$~\cite{Scolnic:2021amr, Brout:2022vxf}. The Union3 compilation, which consists of 2087 SN Ia, many (1363 SN Ia) in common with PantheonPlus~\cite{Rubin:2023jdq}. The DES year 5 (Y5) compilation, which consists of a homogeneously selected sample of 1635 photometrically-classified SN Ia, with redshifts $0.1 < z < 1.3$, that is complemented by 194 low-redshift SN Ia (in common with the PantheonPlus sample) spanning $0.025 < z < 0.1$~\cite{DES:2024jxu}. The {\tt CAMB}-MCMC likelihood analysis again shows evidence for
a time-evolving dark energy equation of state, yielding the following marginalized posterior results:
\begin{equation}
  \left. \begin{array}{ccl}
           w_0 & = & -0.827 \pm 0.063\\
           w_a & = & -0.75^{+0.29}_{-0.25} \\
           \end{array} \right\} \begin{array}{l}{\rm DESI+CMB}\\{\rm +
                                  PantheonPlus}\\
                                  \end{array}\,,
       \end{equation}

       \begin{equation}
  \left. \begin{array}{ccl}
           w_0 & = & -0.64 \pm 0.11\\
           w_a & = & -1.27^{+0.40}_{-0.34} \\
         \end{array} \right\}
\begin{array}{l}{\rm DESI+CMB}\\{\rm +
                                  Union3}\\
                                  \end{array} \,,
       \end{equation}
       
\begin{equation}
  \left. \begin{array}{ccl}
           w_0 & = & -0.727 \pm 0.067\\
           w_a & = & -1.05^{+0.31}_{-0.27} \\
         \end{array} \right\}
\begin{array}{l}{\rm DESI+CMB}\\{\rm +
                                  DESY5}\\
                                  \end{array} \, .
\end{equation}
Table~\ref{tabla1} shows the significance level at which each dataset compilation rejects $\Lambda$CDM. The preference for dynamical dark energy remains robust
when confronting DESI DR1 + CMB + SN Ia datasets with other well-known parameterizations of
$w(z)$~\cite{Giare:2024gpk}.

\subsection{DESI DR2}

In 2025, the DESI Collaboration reported the analysis of DR2~\cite{DESI:2025zgx}. The
constraints on the ($w_0,w_a$) parameters from the MCMC likelihood framework using DESI DR2 BAO data alone are also rather weak,
\begin{equation}
  \left. \begin{array}{ccl}
           w_0 & = & -0.48^{+0.35}_{-0.17}\\
           w_a & < & -1.34 \\
           \end{array} \right\} {\rm DESI \ BAO} \,,
\end{equation}
but they define a degeneracy direction in the
$(w_0,w_a)$ plane, though these constraints do not show a strong preference for
dark energy evolution.

The parameters inferred via {\tt CAMB}-MCMC likelihood analysis using DESI DR2 combined with CMB data show a $3.1\sigma$ preference for evolving dark energy, with best fit values
\begin{equation}
  \left. \begin{array}{ccl}
           w_0 & = & -0.42\pm 0.21 \\
           w_a & = & -1.75 \pm 0.58 \\
           \end{array} \right\} {\rm DESI \ BAO + CMB} \, .
\end{equation}

Finally, {\tt CAMB}-MCMC-derived estimates, produced using DR2 + CMB + SN Ia yield
\begin{equation}
  \left. \begin{array}{ccl}
           w_0 & = & -0.838 \pm 0.055\\
           w_a & = & -0.62^{+0.22}_{-0.19} \\
           \end{array} \right\} \begin{array}{l}{\rm DESI+CMB}\\{\rm +
                                  PantheonPlus}\\
                                  \end{array}\,,
       \end{equation}

       \begin{equation}
  \left. \begin{array}{ccl}
           w_0 & = & -0.667 \pm 0.088\\
           w_a & = & -1.09^{+0.31}_{-0.27} \\
         \end{array} \right\}
\begin{array}{l}{\rm DESI+CMB}\\{\rm +
                                  Union3}\\
                                  \end{array} \,,
       \end{equation}
       
\begin{equation}
  \left. \begin{array}{ccl}
           w_0 & = & -0.752 \pm 0.057\\
           w_a & = & -0.89^{+0.23}_{-0.20} \\
         \end{array} \right\}
\begin{array}{l}{\rm DESI+CMB}\\{\rm +
                                  DESY5}\\
                                  \end{array} \, .
\end{equation}
The $\Lambda$CDM rejection significance for each dataset is detailed in Table~\ref{tabla1}. A point worth noting at this juncture is that rather than choosing between or averaging the three distinct significance levels, the analysis of~\cite{Cortes:2025joz} demonstrates that a proper statistical combination results in a $3.1\sigma$
exclusion. Given these competing results, the most robust inference from the analysis of the DESI Collaboration is the $3.1 \sigma$ exclusion of $\Lambda$CDM obtained by combining DESI and CMB data alone, excluding SN Ia measurements. The preference for dynamical dark energy remains robust
when confronting DESI DR2 + CMB + Union3 with other well-known parameterizations of
$w(z)$~\cite{Shlivko:2025fgv} and when CMB data from the South Pole
Telescope (SPT) is included in the MCMC likelihood analysis~\cite{Li:2025vuh}.

\subsection{Recalibration of SN Ia Datasets}

Recent developments in SN cosmology  have focused on improving the calibration and reducing systematic uncertainties between major datasets. Key updates include the transition to Union3.1, which incorporates updated host galaxy stellar masses, and the Pantheon+ analysis, which
addresses systematic differences in host galaxy stellar mass estimates at $z< 0.15$~\cite{Rubin:2026qdt}. 

Following recalibration, the MCMC likelihood of the DESI DR2 + CMB + SN Ia datasets yields
\begin{equation}
  \left. \begin{array}{ccl}
           w_0 & = & -0.813 \pm 0.055\\
           w_a & = & -0.68^{+0.22}_{-0.19} \\
           \end{array} \right\} \begin{array}{l}{\rm DESI+CMB}\\{\rm +
                                  PantheonPlus \ (recal)}\\
                                  \end{array}\,,
       \end{equation}
favoring evolving dark energy at the $3.2\sigma$ level, and 
        \begin{equation}
  \left. \begin{array}{ccl}
           w_0 & = & -0.719 \pm 0.084\\
           w_a & = & -0.95^{+0.29}_{-0.26} \\
         \end{array} \right\}
\begin{array}{l}{\rm DESI+CMB}\\{\rm +
                                  Union3.1}\\
                                  \end{array} \,,
       \end{equation}
favoring dynamical dark energy at the $3.3\sigma$ level~\cite{Hoyt:2026fve}.

The recalibration of the DESY5 Ia sample, often referred to as the DES-Dovekie reanalysis~\cite{DES:2025sig}, is a critical update aimed at addressing systematics in DESI DR2 cosmological constraints. By utilizing the updated DES-Dovekie calibration, the previously high significance of $4.2\sigma$ for dynamical dark energy, 
 found by combining DESI DR2 with CMB and the original DES5Y data sample is reduced to $3.2\sigma$, with 
\begin{equation}
  \left. \begin{array}{ccl}
           w_0 & = & -0.803 \pm 0.054\\
           w_a & = & -0.72 \pm 0.21 \\
         \end{array} \right\}
\begin{array}{l}{\rm DESI+CMB}\\{\rm +
                                  DES\text{-}Dovekie}\\
                                  \end{array} \, .
\end{equation}

Taken together, these findings indicate that the SN Ia recalibration improves consistency across datasets, leading to a stronger, unified consensus against the $\Lambda$CDM model.

\subsection{DES Validates DESI’s  Evolving Dark Energy Evidence}

DESY6 BAO + DESY5 SN Ia + CMB data yield a DESI-independent
$w_{0}w_{a}$
measurement, with  a $3.2\sigma$ deviation
from $\Lambda$CDM~\cite{DES:2025upx}.
The best fit parameters are
\begin{equation}
  \left. \begin{array}{ccl}
           w_0 & = & -0.673^{+0.098}_{-0.097} \\
           w_a & = & -1.37^{+0.51}_{-0.50} \\
         \end{array} \right\}
\begin{array}{l}{\rm DESY6\text{-}BAO+CMB}\\{\rm +
                                  DESY5}\\
                                  \end{array} \, .
\end{equation}
The 
significance decreases to $2.4 \sigma$ when adopting the DES-SN
Dovekie recalibration~\cite{DES:2026aht}. For a recent summary of DES-SN
results, see~\cite{DES:2026jmi}. Multiple data set analyses
in~\cite{Ishak:2025cay,Giare:2025pzu,Giare:2024oil} show comparable findings.

\subsection{Navigating the Nuances of DESI’s Dark Energy Analyses}

The DESI+CMB+SN Ia constraints have an
unambiguous preference for a sector of the $(w_0,w_a)$ plane in which
$w_0 > -1$ and $w_0 + w_a < -1$, suggesting that $w(z)$ may have experienced a transition from a phase violating the null
energy condition at large $z$ to a phase obeying it at small
$z$. As shown in~\cite{Shlivko:2024llw},  this impression may be
misleading, because rather simple quintessence models satisfying the
null energy condition for all $z$, characterized by hilltop potentials
with a sharp decline are compatible with DESI data. Along this line,
it was noted in~\cite{Abreu:2025zng} that while DESI data prefer
$(w_0,w_a)$ to one-parameter characterization of scalar-field models,
the SN Ia data prefer a scalar field to $(w_0,w_a)$, and as previously
noted in~\cite{DESI:2025zgx}, together they favor a $(w_0,w_a)$
model. Furthermore, combining standard DESI DR2, CMB, and SN Ia
datasets with large-scale structure and strong lensing measurements
(TDCOSMO + SLACS) provides additional support for  scalar field
models~\cite{Shajib:2025tpd}.

An integrated analysis of DESI DR1 with CMB observations under the $\Lambda$CDM
framework has established a stringent (95\%~CL) upper limit on the
total neutrino mass of $\sum m_{\nu} \leq 0.072~{\rm eV}$~\cite{DESI:2024mwx}. This
constraint is nearing the absolute minimum masses established by
neutrino oscillation experiments: $\sum
m_{\nu} \geq 0.057/0.10~{\rm eV}$ for normal/inverted
hierarchy~\cite{Esteban:2024eli}. Escalating the situation, when the strict physical prior
$\sum m_\nu > 0$ is relaxed, the data actually suggest a statistical
preference for a negative neutrino mass
sum~\cite{Loverde:2024nfi,Craig:2024tky,Green:2024xbb}. Actually,
  CMB-only contours exhibit a positive correlation between $\sum
  m_\nu$ and $\Omega_{\rm CDM} h^2$ that is partially broken when BAO
  distance measurements and SN Ia data are included in the
  analysis~\cite{Loverde:2024nfi}. Therefore, a negative $\sum
m_{\nu }$  reflects either a deficit in late-time matter abundance
relative to the well-constrained $\Omega_m$ fiducial value (measured
to 1\% precision), or a fundamental tension between standard neutrino
physics and the $\Lambda$CDM framework. Overall, this inconsistency highlights potential limitations in the
$\Lambda$CDM model, encouraging exploration into alternative
cosmologies, particularly dynamical dark energy, to resolve inherent
degeneracies~\cite{Craig:2024tky,Green:2024xbb,Elbers:2024sha,Jiang:2024viw}.\footnote{A degeneracy between the
  effects of neutrino masses and baryon-dark matter interactions on
  cosmological observables could also weaken constraints on neutrino   
  masses~\cite{Anchordoqui:2025elg}.}
By adopting the $w_0w_a$CDM parameterization or a matter-to-dark energy conversion mechanism, the tight restrictions on neutrino mass imposed by DESI data are relaxed, enabling positive neutrino masses that are more consistent with both
oscillation lower bounds and cosmological
data~\cite{Elbers:2025vlz,DESI:2025ffm,Du:2025xes,Yang:2026yaq}. For example, the upper bound
on neutrino mass assuming the CPL parametrization relaxes
significantly, typically to around $\sum m_{\nu} < 0.163~{\rm
  eV}$~\cite{Elbers:2025vlz}. Now, the BAO-CMB-SN Ia incompatibility
with neutrino masses is driven by the matter-era distance
interval between photon-baryon decoupling and the redshifts probed by
DESI~\cite{Weiner:2026sfm}. This finding critically impacts how we interpret preferences for
the $w_0$-$w_a$ parametrization; particularly low-redshift dark
energy dynamics alone cannot resolve the tension. Instead, as
demonstrated analytically and numerically in~\cite{Weiner:2026sfm}, the solution requires
the dark energy density to be substantially suppressed at redshifts
strictly above $z > 2.33$. This specific resolution to the matter-era
distance excess is precisely how $w_{0}$-$w_{a}$ relaxes neutrino mass
bounds~\cite{Costa:2025kwt}. One final comment, neutrino mass limits
from DESI DR2 are consistent regardless of which SN Ia dataset is used (including
DES-Dovekie)~\cite{Yang:2026yaq}. This stability persists even when using only DESI and CMB data without any supernova input at all.

In closing, we note that while many interpret DESI DR2 as evidence for
dynamical dark energy, follow-up studies suggest this finding may
depend heavily on dataset combinations, SN Ia calibration, and
model-comparison methods, with some Bayesian analyses showing a
weakened or absent
preference~\cite{Patel:2024odo,Efstathiou:2024xcq,Colgain:2024mtg,Colgain:2025nzf,Wang:2025bkk,Efstathiou:2025tie,Ong:2025utx,Ong:2026tta}. In
particular, Bayesian analyses favoring the $\Lambda$CDM model indicate
that support for dynamical dark energy stems mainly from reconciling
dataset discrepancies~\cite{Ong:2025utx,Ong:2026tta}. Recently, by
enhancing cross-calibration, the DES-Dovekie, Union3.1, and
PantheonPlus samples mitigated key systematic uncertainties related to
combining SN Ia
datasets~\cite{DES:2025sig,Hoyt:2026fve,Rubin:2026qdt}. Nevertheless,
the elephant in the room is the $H_0$ tension, a major cosmological
crisis stemming from a $\sim 10\%$ discrepancy in the measured rate of
the universe's expansion~\cite{Abdalla:2022yfr}. Low-$z$ measurements
reported e.g., by the Local Distance Network ($H_0$DN) yield a higher
value ($0.73< h < 0.74$)~\cite{H0DN:2025lyy} compared to high-$z$
predictions ($0.67 < h < 0.68$) based on the Planck satellite's CMB
data calibrated using $\Lambda$CDM~\cite{Planck:2018vyg}. At present,
there is roughly a $7\sigma$ tension between the late and early universe determination of $H_0$~\cite{CosmoVerseNetwork:2025alb}.
The inferred values of $H_0$ as reported in~\cite{DESI:2025zgx} using
possible DR2 combination datasets are: {\it (i)}~DESI DR2 + CMB, $h =
0.636^{+0.016}_{-0.021}$; {\it (ii)}~DESI DR2 + CMB + PantheonPlus,
$h = 0.6751\pm 0.0059$; {\it (iii)}~DESI + CMB + Union3, $h = 0.6591
\pm 0.0084$; {\it (iv)} DESI+CMB+DESY5, $h = 0.6674 \pm 0.0056$. As a
sharp reader might have noticed, incorporating the $H_0$ prior into
CMB, DESI DR2 and Pantheon Plus/Union3/DESY5 compilation dataset
reduces the preference for dynamical dark energy to
$1.5\sigma$/$1.4\sigma$/$2.4\sigma$ level,
respectively~\cite{Pang:2025lvh}. On the flip side, analyzing local distance data with the $w_0w_a$ dark
energy models favored by DESI yields a smaller value of $H_0$ than the
value obtained assuming $\Lambda$CDM~\cite{Turner:2026zlj}. This decrease can be as large as
$2.5~{\rm km \, s}^{-1}\, {\rm Mpc}^{-1}$using local measurements
alone. However, when adding external constraints, the downward shift
shrinks to $1.1 \pm 0.38~{\rm km \, s}^{-1}\, {\rm Mpc}^{-1}$ for DESI
+ CMB data, and drops further to $0.5 \pm 0.1~{\rm km
  s}^{-1}~{\rm Mpc}^{-1}$ when combining DESI, CMB, and SN Ia data.

Notwithstanding the profound challenges posed by the $H_0$ tension,
the strong  evidence against $\Lambda$CDM indicated in
Table~\ref{tabla1} and the comparative analysis
of~\cite{Zhang:2025lam} make a compelling case for the investigation
of string-inspired quintessence alternatives.  In Sec.~\ref{sec:4} we review stringy quintessence models that are consistent with the
 observational data. After that, in Sec.~\ref{sec:5} we survey  stringy scenarios in which a quintessence scalar field (with
positive kinetic energy) couples to dark matter. This type of models aligns with the region favored by DESI results, featuring a physically well-behaved dark energy sector ($w_\phi > -1,$ $\forall
\phi$) that allows the effective equation of state $w_{\rm eff}$ to evolve from a {\it phantom regime} $w_{\rm eff} < -1$~\cite{Caldwell:2003vq} in the distant past into the non-phantom regime $w_{\rm eff} > -1$ in the present epoch.

\section{Cosmological implications  of Swampland Conjectures}
  \label{sec:3}

While both the Standard Model (SM) of particle physics and 
$\Lambda$CDM are empirically accurate, they rely on unnatural fine-tuning, leaving them without a simple theoretical explanation. Indeed, these frameworks lack a clear explanation for why the weak scale is so small compared to gravity (gauge hierarchy problem) or why dark energy is so small (cosmological hierarchy problem), forcing us to confront puzzling coincidences in nature. The traditional understanding of naturalness, which relies on symmetries to decouple low-energy (IR) physics from high-energy (UV) details, appears to fail in explaining these hierarchies. Such a failure suggests a need for a new theoretical paradigm, as the assumption that UV physics is irrelevant to the IR seems to be  incorrect. As a matter of fact, string theory teaches us that UV and IR physics are often inextricably mixed. A prime example of this UV/IR mixing is found in black holes: they are IR objects described by Einstein's equations, yet their Bekenstein-Hawking entropy~\cite{Bekenstein:1973ur,Hawking:1974rv} shows that low-energy physics encodes information about high-energy quantum states~\cite{Vafa:2024fpx}. This implies that the conventional approach to naturalness fails precisely because it ignores this mixing and fails to account for quantum gravity, which is typically dismissed as irrelevant at low energies. A robust notion of naturalness must incorporate constraints from the UV completion of quantum gravity, as UV consistency conditions heavily restrict the permissible IR physics. The swampland conjectures, which incorporate these gravity-induced constraints, may alleviate, or at least better explain, some of the fine-tuning problems in our current physical theories.

In this section, we review key swampland conjectures, offer a possible explanation for the cosmological hierarchy problem, justify why $\Lambda$CDM is considered part of the swampland (i.e., lacks a consistent UV completion in quantum gravity), and derive constraints for quintessence-field potentials. Before moving forward, we define the genetic framework and introduce the Lagrangian density of our physical system.

In extra-dimensional physics, if you have a compact internal dimension, its size $L$ inversely dictates the mass of the resulting particles: the larger the dimension, the lighter the {\it tower} of states. These are known as Kaluza-Klein (KK) particles, and they can be understood as higher-dimensional particles whose momentum within that extra dimension manifests to us as mass,
  $m_{\rm KK} \sim 1/L$~\cite{Kaluza:1921tu,Klein:1926tv}. The mass spectrum is often indexed by an integer $n$ (often called the KK index or KK number), $m_{{\rm KK},n} \sim n/L$. String theory generally implies the existence of a tower of weakly coupled states, often arising from KK particles or string excitations~\cite{Ooguri:2006in,Lee:2019wij}.
Essentially, by reducing the coupling in string theory, which governs the gravitational interaction and sets the string mass scale $M_s$, a tower of light string excitations emerges~\cite{Lee:2018urn}. Therefore, in weak coupling, a tower of states exists with masses proportional to $M_s$. This structure is a fundamental feature related to consistency, often connecting to the {\it weak gravity conjecture} (WGC)~\cite{Arkani-Hamed:2006emk,Harlow:2022ich}, which suggests such states exist to maintain gravitational constraints.

Intuitively, because gravitons couple to all forms of matter and energy, a high number of available states can create loop corrections in the graviton propagator. These corrections can cause gravity to become strongly coupled at a much lower energy scale than the expected $M_p$. In other words, quantum gravity effects are expected to become important at a cut-off species scale 
$\Lambda_s$~\cite{Dvali:2007hz,Dvali:2007wp,Dvali:2009ks,Dvali:2012uq}
(for some more recent work 
see~\cite{Castellano:2022bvr,vandeHeisteeg:2022btw,Cribiori:2022nke,Cribiori:2023ffn,Cribiori:2023sch,Calderon-Infante:2023ler,Calderon-Infante:2023uhz,vandeHeisteeg:2023dlw,Castellano:2023aum,Bedroya:2024uva,Calderon-Infante:2025pls}, for a recent review see~\cite{Ibanez:2026yyk}),
 which can be much lower than $M_p$ whenever one has a large
 number of particle species becoming light. Concretely the species scale and the number of light species $N_s$ are related as follows~\cite{Dvali:2007hz}
 \begin{equation}
 \Lambda_s =
 M_{p,d}/ N_s^{1/(d-2)}\, ,
\label{species}
\end{equation}
where $d$ is
the number of spacetime-dimensions of the low-energy effective field
theory and $M_{p,d}$ is the reduced Planck mass in $d$ dimensions. The limit of a large number of species  is predicted to
 happen at any perturbative limit of an effective field theory coupled
 to gravity, or equivalently in any infinite distance limit within the
 field space of the quantum gravity completion~\cite{Ooguri:2006in}.

Now, consider a reduction from a higher-dimensional $D$-theory to a lower-dimensional $d$-theory. In the presence of gravity, the internal space volume ${\cal V}$ is determined by a scalar field (modulus) $\phi$. Following canonical normalization of the kinetic term, $\mathscr{L} \supset (\partial \phi)^2/2$, the volume dependence on a scalar field is an exponential function of the field, with a coefficient related to the dimensionality of the reduction 
 \begin{equation}
   {\cal V} \sim e^{\sqrt{(D-2) (D-d)/(d-2)} \phi} \, ,
\end{equation}
and the mass of the associated KK particles decreases exponentially
\begin{equation}
m_{\rm KK} \sim  e^{-\phi \sqrt{(D-2) (D-d)/(d-2)}} \, ;
\label{mDC1}
\end{equation}
for further details, see e.g. Sec.~5.5 of~\cite{Agmon:2022thq}. In string theory, the dilaton is a scalar field $\phi$ that determines the string coupling constant (interaction strength) and influences string tension. While not universal for all string types, excitation masses in certain models show an exponential dependence on this field. Notably, string states involving gravitons always demonstrate this exponential behavior,
\begin{equation}
M_s \sim e^{- \phi/\sqrt{d-2}} \, ;
\label{mDC2}
\end{equation}
for further details, see e.g. Sec.~5.5 of~\cite{Agmon:2022thq}. Gravity seems to be the {\it missing link} in this pattern. Equations~(\ref{mDC1}) and (\ref{mDC2}) set the stage for the {\it distance conjecture} (DC), which posits that pushing a scalar field toward infinity inevitably triggers a tower of states that becomes increasingly light~\cite{Ooguri:2006in}. 
This behavior makes weak-coupling limits increasingly controllable. In fact, approaching these limits reveals a previously hidden, weak-coupling structure that offers a novel theoretical framework that departs from conventional naturalness. Rather than a single scalar field, this framework allows for multiple fields, meaning the system can depend on a multi-dimensional scalar field space, known as the moduli space.
This field space has its own metric derived from the kinetic terms in the action. As the careful reader may have inferred,  the limit of a large number of species coincides with the asymptotic limits in moduli space.

Our starting point is then the low-energy EFT of some
scalar fields (moduli) that are coupled to gravity,
\begin{eqnarray}
S_{\rm EFT}&=&\int d^dx
               \sqrt{-g}\biggl\{M_{p,d}^{d-2}\left[\frac{1}{2} \
               {\cal R} - \frac{1} {2} {g^{\mu\nu} \ G_{ij} \
               \partial_\mu \phi^i \, \partial_\nu \phi^j} - V(\phi) +
               \mathscr{L}_{m,r} \right.               \nonumber\\
           &+ & \left.
\sum_{n > 2} \Lambda_s^{2-n}(\phi){\cal O}_n({\cal R})\right]
+ \sum_{n>2 }m_{\rm lightest}^{d-n}(\phi) \ {\cal O}_n({\cal R})+\dots\biggr\}
 \,,\label{eft}
\end{eqnarray}
where ${\cal
  R}$ is the Ricci scalar, $g$ refers to the space-time metric, the
scalar fields $\phi^i$ correspond to the moduli of the string
compactification, $G_{ij}$ is the moduli space metric, $V(\phi)$ a
possible moduli dependent scalar potential, and $\mathscr{L}_{m,r}$ is
the matter (cold dark and baryonic) and radiation Lagrangian density. In addition,
the higher order corrections appearing in the action can be written as
a double EFT expansion~\cite{Calderon-Infante:2025ldq} on operator-valued functions ${\cal O}_n ({\cal R})$ suppressed by the moduli dependent species scale
$\Lambda_s(\phi)$ and the characteristic mass-scale of the lightest
tower of states $m_{\rm lightest}(\phi)$~\cite{vandeHeisteeg:2022btw,
  Bedroya:2024uva}. In this context, ${\cal O}_n({\cal R})$ is a
schematic notation representing an operator-valued function of dimension
$n$,  built purely out of the Riemann tensor, not necessarily the Ricci scalar raised to the power of $n$.

With this in view, we now proceed to discuss the cosmological implications of swampland conjectures.

\subsection{The Cosmological Hierarchy Problem and the Dark Dimension}
\label{sec:3a}

To figure out how a small $\Lambda$ can be reconciled with a UV complete
description of quantum gravity we are interested in the swampland
conjectures that place constraints on the cosmological dynamics of a
scalar field $\phi$ that rolls down the potential $V(\phi)$.

We have seen that the DC constrains field excursions to be small over cosmic history,
\begin{equation}
  \frac{\Delta \phi}{M_p} \equiv c \lesssim {\cal O}(1) \, ,
\label{distancecon}
\end{equation}
because when venturing to large distances within scalar field space of any consistent theory of quantum gravity,
a tower of particles will become light at a rate that is exponential
in the field space distance~\cite{Ooguri:2006in}.\footnote{For a teaching-focused discussion, see~\cite{vanBeest:2021lhn}.}  The DC relates the tower mass scale $m$ to the moduli space metric $G_{ij}$ in the following way:
\begin{equation}
m (\phi) \sim e^{-\alpha\, \Delta \phi}\, ,
\label{mfollowing}
\end{equation}
where distance and masses are measured in Planck units, $\alpha$ is an order one positive constant, and $\Delta \phi$ is the geodesic distance from an arbitrary reference point $\phi_0$ to the point labeled by $\phi$, for asymptotically
large $\phi$. This distance is computed from the moduli space metric as 
\begin{equation}
  \Delta \phi = \int_0^1 d\lambda \ \sqrt{{d\phi^i\over d\lambda} \
    G_{ij}(\lambda) \ {d\phi^j\over d\lambda}}\ ,
\end{equation}
where $\lambda$ is a parameter of the path, with $\lambda =0$
corresponding to $\phi_0$ and $\lambda = 1$ to $\phi$. 

Associated to the DC
is the {\it anti-de Sitter} (AdS) {\it distance conjecture} (AdS-DC),
which states that in the asymptotic limit of a small cosmological
constant $|\Lambda| \rightarrow 0$, there is always a light tower
of states of typical mass scale $m$, which scales as
\begin{equation}
m \sim |\Lambda|^{1/\mathtt{a}}\,  ,
\label{AdS-DC}
\end{equation}
with $\mathtt{a}$ an order one  positive
constant~\cite{Lust:2019zwm}. In this context, the parameter $\Lambda$ characterizes the AdS metric field space, in which the limit $|\Lambda| \to 0$ lies at an infinite distance. Actually, if the scaling behaviour remains valid in de Sitter (dS) space, an unbounded number of massless modes would also emerge in the
limit $|\Lambda| \to 0$. Herein, we assume that the AdS-DC also holds
for positive vacuum energy. 

Since the KK tower contains massive spin-2 bosons, there is a strong constraint from fundamental physics, {\it unitarity}, which is expressed in the form of the Higuchi bound $\mathtt{a} \geq 2$ and imposes an absolute upper limit on $1/\mathtt{a}$~\cite{Higuchi:1986py}. Besides, $1/\mathtt{a}$ has a lower limit set by contributions of the Casimir energy; in four dimensions $\mathtt{a} \leq 4$~\cite{Montero:2022prj}. A theoretical amendment on the connection between the cosmological and KK mass scales confirms $\mathtt{a} = 4$~\cite{Anchordoqui:2023laz}. 

Actually, the AdS-DC has paved the ground for a possible explanation of
the bafflingly small value of dark energy, $\Lambda \sim 7 \times  10^{-121} M_p^4 \sim (2~{\rm meV})^4$, by linking it to the existence of a single mesoscopic extra dimension of size $R$, dubbed the dark dimension~\cite{Montero:2022prj}.\footnote{It is worth noting that the cosmological constant, $\Lambda$, bridges two distinct length scales: the large-scale curvature of the observable Universe ($[\Lambda] = L^{-2}$) and the much smaller characteristic length scale of dark energy ($L^{-4}$), derived from [$8\pi M_p^2 \Lambda$]. Throughout this review, we identify both quantities simply as $\Lambda$, consistent with our frequent use of Planck units.} Indeed, taking $\mathtt{a}=4$, we arrive at the relation $\Lambda(R) \sim 1/R^4$. Whence, it is tempting to speculate whether we could be living near an asymptotic limit of moduli space, in which $R$ is large and $\Lambda$ very small in Planck units.
More concretely, tf the dark dimension  scale $R$
reaches the limits of current Newtonian gravity experiments, which
have confirmed the inverse-square law down to approximately 30 or $40~\mu{\rm m}$~\cite{Lee:2020zjt,Tan:2020vpf}, the resulting cosmological constant aligns with the observed density of dark energy.

Within the dark dimension scenario the SM fields are localized on a 3-brane transverse to the internal dimension, while gravity spills
into the bulk of the compact
space~\cite{Arkani-Hamed:1998jmv,Antoniadis:1998ig}. Alternatively,
the dark dimension can be understood as a line interval with
end-of-the-world 9-branes attached at each
end~\cite{Schwarz:2024tet}. Of course, this is equivalent to a
semicircular dimension endowed with $S^1/\mathbb{Z}_2$ symmetry.

Highly sensitive, short-range gravity tests~\cite{Westphal:2020okx} could
  potentially distinguish between these two geometries~\cite{Schwarz:2024tet}. For a circular
  extra dimension of radius $R$, the summed Yukawa potentials from KK gravitons yield a potential proportional to
\begin{equation}
V(r) \propto \frac{1}{r} \sum_{n=-\infty}^{\infty} e^{-|n| \, r/R} =
\frac{1}{r} \, \frac{1+e^{- r/R}}{1-e^{-r/R}} \sim
\frac{1}{r} \left(1+ 2e^{-r/R} +\ldots \right) \, .
\end{equation}
Conversely, an orbifold/line segment configuration results in
\begin{equation}
V(r) \propto \frac{1}{r} \ \sum_{n=0}^{\infty} e^{-|n| \, r/R} = 
\frac{1}{r} \,
 \frac{1}{1-e^{- r/R}} \sim \frac{1}{r} \left(1+
e^{-r/R} +\ldots \right) \, .
\end{equation}
While both cases exhibit the expected $R/r^2$ behavior at small distances, they differ by a factor of two.

Table-top experiments probing the gravitational inverse-square law
typically constrain a supplementary Yukawa interaction, in which the predicted potential energy between two masses, $m_{1}$ and $m_{2}$, separated by a distance $r$ takes the form
\begin{equation}
  V(r) = - G \frac{m_1 m_2}{r}
  \left( 1 + |\alpha_s| \, e^{-r/\lambda_s} \right) \,,
\end{equation}
where $G$ is Newton's constant, $|\alpha_s|$ is the strength of the new
potential as compared to the Newtonian gravitational
potential, and $\lambda_s$ is its
range~\cite{Antoniadis:1997zg}. It is easily seen in Fig.~5 of~\cite{Lee:2020zjt} and
Fig.~6 of~\cite{Tan:2020vpf} that the 95\%~CL upper limit on $\lambda _{s}$ is approximately
$30~\mu{\rm m}$ for $|\alpha_s| = 2$ and $40~\mu{\rm m}$ for
$|\alpha_s| = 1$. These bounds imply characteristic mass scales of
$m_{\rm KK} \simeq 6.6~{\rm meV}$ and $m_{\rm KK} \simeq 4.9~{\rm
meV}$, respectively. In summary, if the dark dimension has a 10-micron-scale characteristic length, then the
KK graviton tower necessarily opens up at the mass scale
\begin{equation}
  m_{\rm KK} \sim 1/R \sim \Lambda^{1/4} \sim {\cal O}({\rm meV})
\end{equation}
and the species scale is estimated to
be~\cite{Arkani-Hamed:1998jmv}
\begin{equation}
  \Lambda_s \sim m_{\rm KK}^{1/3} \   M_p^{2/3} \sim 10^{8.6}~{\rm GeV} \, .
\end{equation}  
We note in passing that there is borderline experimental feasibility of having two dark dimensions of micron scale~\cite{Anchordoqui:2025nmb}. Cosmology provides the most critical bounds~\cite{Arkani-Hamed:1998sfv,Hall:1999mk}.

Beyond offering a potential solution to the cosmological hierarchy problem, the 
 dark dimension provides a colosseum for dark matter contenders, including the decaying tower of massive KK gravitons~\cite{Gonzalo:2022jac,Anchordoqui:2022svl,Law-Smith:2023czn,Obied:2023clp}, whose evolution shapes the dynamical dark matter framework~\cite{Dienes:2011ja}, and 5D primordial black holes in the asteroid mass window~\cite{Anchordoqui:2022txe,Anchordoqui:2022tgp,Anchordoqui:2024akj,Anchordoqui:2024dxu,Anchordoqui:2024jkn,Anchordoqui:2024tdj,Anchordoqui:2025xug,Ettengruber:2025kzw,Anchordoqui:2025opy,Leontaris:2025piz,Leontaris:2026kvu}. The dark dimension also offers a novel approach to the study of supersymmetry breaking~\cite{Anchordoqui:2023oqm}, 
axion physics~\cite{Gendler:2024gdo}, and neutrino masses~\cite{Anchordoqui:2023wkm,Anchordoqui:2024xvl,Antoniadis:2025rck,Montero:2025hye,Bai:2026kdq}.\footnote{The idea that the smallness of the neutrino mass might be ascribed to the fact that
right-handed neutrinos could live in the
bulk was introduced in~\cite{Dienes:1998sb,Arkani-Hamed:1998wuz,Dvali:1999cn}. The
coupling of right-neutrinos to the left-handed SM neutrinos living on the brane is
inversely proportional to the square-root of the bulk volume.}
Attempts at string constructions of the dark dimension have been
explored in~\cite{Blumenhagen:2022zzw,Dudas:2025yqm,Blumenhagen:2026rgu} and worldsheet 
aspects of the dark dimension were discussed in~\cite{Basile:2024lcz}.

\subsection{Why $\Lambda$CDM is Banished to the Swampland}

\label{sec:3b}

Construction of dS vacua in the string landscape has proven to be
harder than expected, and actually no attempt has been fully
successful so far; for an educational breakdown of the subject, see~\cite{VanRiet:2023pnx}. In response to this obstructive fact the {\it dS
  conjecture} (dSC) places restrictions on the
scalar field potential, which must obey either 
\begin{equation}
  M_p \frac{|\nabla V|}{V} \equiv c' \gtrsim {\cal O} (1)
\label{dSC}
\end{equation}
or else
\begin{equation}
  - M_p^2 \frac{{\rm min} (\nabla_i \nabla_jV )}{V} \equiv c'' \gtrsim {\cal O} (1) \, ,
  \label{RdSC}
\end{equation}
where the operator $\nabla$
 is the Levi-Civita connection associated with the moduli space metric 
$G_{ij}$ and ${\rm min} (\nabla_i \nabla_jV )$
 stands for the minimum eigenvalue of the
Hessian $\nabla_i \nabla_j V$  in an orthonormal
frame~\cite{Obied:2018sgi,Ooguri:2018wrx}. A very similar
version of the bound (\ref{dSC}) was initially proposed
in~\cite{Dvali:2014gua,Dvali:2017eba}; for a detailed comparison
between the two constraints on the scalar potential,
see~\cite{Dvali:2018fqu,Dvali:2018jhn}. The bounds (\ref{dSC}) and (\ref{RdSC}) imply that
the potential is either steep or concave-down hill, respectively. The condition (\ref{RdSC}) is applied near
the maximum of the potential, if one exists. It is worth noting that these constraints are trivially met if the
potential is non-positive or in the $M_p \to \infty$ limit.

The no-dS conjecture fundamentally conflicts with $\Lambda$CDM
cosmology, as a positive cosmological constant violates the defined
bound  (\ref{dSC})  where $c'>0$~\cite{Agrawal:2018own}. Given the
critical nature of this conflict, further investigation into the dS
conjecture is warranted. Entropy-based arguments can be used to derive the
 dS conjecture (\ref{dSC}), from the distance
conjecture (\ref{distancecon})~\cite{Ooguri:2018wrx}. This derivation
holds in the weak coupling regime of string theory, where physical
observables remain calculable. Specifically, assuming $\phi$ increases
over
time, the distance conjecture implies that a growing number of
particle 
species, $N_s(\phi)$, becomes light enough to be excited and must be
included in the low-energy EFT description. The population of these light
particles, characterized by masses given in (\ref{mfollowing}),
scales as
\begin{equation}
  N_s(\phi)  \sim  n(\phi) \ \exp\left(\beta
    \frac{\phi}{M_p}\right) \label{speciesphi} \,, 
\end{equation} 
where $\beta$ is a positive constant determined by the string tower properties
and $n(\phi)$ is a monotonically increasing function such that: 
\begin{equation}
 \frac{dn(\phi)}{d\phi}  >   0 \, .
\label{dndphi}
\end{equation}

In an accelerating universe with a Hubble horizon ${\cal R}_H=
H^{-1}$, the growing number of degrees of freedom
(\ref{speciesphi}) drives an
increase in the entropy of the tower of states, parameterized by
\begin{equation}
{\cal S}_{\rm tower} \left(N_s,{\cal R}_{H}\right) = N_s^{\gamma} \ \left(M_p {\cal
  R}_{H} \right)^{\delta} \,,
\end{equation}
where $\gamma >0$ and $\delta\geq 0$ are constants (e.g., $\delta=0$ for
point-particle behavior).\footnote{Adopting a holographic perspective,
  we express the Hubble rate $H$ by setting the horizon radius to be
  roughly the inverse of the Hubble parameter, see
  e.g.~\cite{Myung:2005sv}.} According to~\cite{Bousso:1999xy}, this
tower entropy is constrained by the Gibbons-Hawking
entropy~\cite{Gibbons:1977mu}, satisfying:
\begin{equation}
N_s^{\gamma }(M_p{\cal R}_H)^{\delta }\leq 8\pi ^{2}{\cal R}_H^{2}
M_{p}^{2} \, .
\label{Bousso_ineq}
\end{equation}

By assuming the total energy is dominated by potential energy $V$, it follows that
\begin{equation}
H^2 \sim \frac{|V|}{3M_p^2} \,,
\label{HsqrtVrealtion}
\end{equation}
as
detailed e.g. in the Appendix of~\cite{Anchordoqui:2026nit}.
Accordingly, the Hubble rate allows us to rewrite (\ref{Bousso_ineq}) as
\begin{equation}  \frac{V}{3M_p^4} \, \leq \, \left( \frac{8
      \pi^2}{N_s^{\gamma}} \right)^{1/(1 - \delta/2)}
  . \label{VN} \end{equation}
Taking logarithm and
differentiating with respect to $\phi$ gives
\begin{equation} 
  \frac{V'}{V} \leq  - \frac{\gamma}{1 - \delta/2} \ \left(\log N_s
  \right)'  \,,
\label{Vprimeneg}
\end{equation}
where primes indicate
$\phi$-derivatives. Note that the field $\phi$ rolls towards larger values in
a potential that decreases. Concurrently, $(\log N_s)' > 0$  because the number of light species must increase exponentially as the field moves into the large-distance regime. Since $V'$ is negative,
we can rewrite (\ref{Vprimeneg}) as
\begin{equation}
  \frac{|V'|}{V} \geq \frac{\gamma}{1 - \delta/2} \ (\log N_s)' \,.
\end{equation}
Applying Eqs.~(\ref{speciesphi}) and (\ref{dndphi}) leads to
\begin{equation}
\frac{|V'|}{V}  \geq \frac{2\gamma}{2-\delta} \left(
                              \frac{n'}{n}+\frac{\beta}{M_{p}} \right)
                            >  \frac{2\beta
                              \gamma}{(2-\delta) \ M_p},
\label{Vfinal}
\end{equation}
with $\delta<2$~\cite{Ooguri:2018wrx}. Note that for a single field, $\nabla$
simplifies to $\partial_\phi$, and thus comparing (\ref{Vfinal}) with ~(\ref{dSC}) allows us to identify the constant
$c'$ as
\begin{equation}
  c'=\frac{2\beta\gamma}{2-\delta} \, .
\end{equation}

One final takeaway is that the dSC implies that single-field inflation, characterized
by an exceptionally flat potential ($\nabla V \approx 0$), is
incompatible with the string landscape and belongs in the swampland,
see e.g.~\cite{Achucarro:2018vey,Garg:2018reu,Ben-Dayan:2018mhe,Kinney:2018nny,Fukuda:2018haz,Garg:2018zdg,Agrawal:2018rcg,Chiang:2018lqx,Brandenberger:2020oav}.

\subsection{Censoring the Smallest Scales}
\label{sec:3c}

Cosmic inflation is a widely accepted framework that explains the
uniformity (horizon problem) and geometry (flatness problem) of the
universe by positing a phase of rapid, exponential expansion in the
early
universe~\cite{Guth:1980zm,Starobinsky:1980te,Linde:1983gd}. Despite
its success, the framework encounters a fundamental hurdle known as
the trans-Planckian problem~\cite{Martin:2000xs}. Throughout
inflation, quantum fluctuations are exponentially stretched. If the
inflationary epoch is sufficiently long, the primordial perturbations
we observe in the CMB would have originated at length scales shorter
than the Planck length $\ell_p = M_p^{-1}$. At this sub-Planckian regime, general relativity collapses and quantum gravity dominates. Consequently, predictions of these fluctuations rely on the unverifiable assumption that effective quantum field theory can be extrapolated beyond its physical limits.

The {\it trans-Planckian censorship conjecture} (TCC) attempts to
address the trans-Planckian problem by postulating that fluctuations
with sub-Planckian length scales are strictly forbidden from exiting the Hubble
horizon and freeze, because any
valid theory of quantum gravity will always act as a
censor~\cite{Bedroya:2019snp}. Strictly speaking, the TCC states that in a uniformly expanding universe with scale
factor $a$ and Hubble parameter $H(t) = \dot a/a$, it should not be possible for a sub-Planckian region to
become larger than the Hubble horizon ${\cal R}_H = H^{-1}$; {\it viz.},
\begin{equation}
  \frac{a_f}{a_i} < \frac{{\cal R}_H}{\ell_p} =
  \frac{M_p}{H_f} 
\label{TCC}
\end{equation}
for any initial and final times $t_i < t_f$. Since $\int H(t) \ dt = \log(a_f/a_i)$ it follows that the $e$-fold
number $\Delta N = \log (a_f/a_i)$ should obey
\begin{equation}
  \Delta N = \int_{t_i}^{t_f} H \ dt < \log \frac{M_p}{H_f} \, .
\label{TCCefolds}
\end{equation}
In summary, the TCC states that the universe is {\it censored} from the possibility of observing the chaotic, quantum-scale physics of gravity;
sub-Planckian quantum fluctuations and  all information about
trans-Planckian scales must remain hidden from the {\it classical
  domain}, {\it viz} the super-Hubble domain where
fluctuations grow and {\it can} classicalize.

The censorship rule~(\ref{TCC}) places very restrictive constraints on the very early universe
cosmology~\cite{Bedroya:2019tba,Mizuno:2019bxy,Bedroya:2020rmd,Brandenberger:2021pzy,Vafa:2025nst}. Specifically,
assuming immediate reheating after inflation, (\ref{TCC}) leads to an upper
bound
\begin{equation}
  H_f \equiv H_{\rm inf} \lesssim 10^{-20} M_p,
  \label{Hinfbound}
\end{equation}
which directly implies that the TCC predicts an extremely small
tensor-to-scalar ratio
\begin{equation}
  r = \frac{2}{\pi^2 \ {\cal P}_s}\left(\frac{H_{\rm inf}}{M_p}\right)^2 <
    6.8 \times 10^{-33} \, ,
\label{r}
\end{equation}
for primordial gravitational
waves~\cite{Bedroya:2019tba}.\footnote{The tensor-to-scalar ratio
  quantifies the amplitude of primordial gravitational waves (tensor
  perturbations) relative to density fluctuations (scalar
  perturbations) generated during cosmic inflation, generally defined
  as $r \equiv {\cal P}_t/{\cal P}_s$.} In (\ref{r}) we
have used  the primordial power spectrum of scalar density fluctuations
${\cal P}_s \approx 2.1 \times 10^{-9}$ as reported by the Planck
Collaboration~\cite{Planck:2018vyg}.  During inflation, the universe is dominated by the vacuum energy 
$V$ and thus the Friedmann equation is given by (\ref{HsqrtVrealtion}).  Substituting
(\ref{Hinfbound}) into (\ref{HsqrtVrealtion}) leads an upper bound for
the energy scale of inflation
\begin{equation}
\eta = V^{1/4}= \left(3H_{\rm inf}^{2}M_{p}^{2} \right)^{1/4} \lesssim 10^9~{\rm GeV}
\, .
\end{equation}
Relaxing the assumption of instantaneous reheating post-inflation
eases the constraint (\ref{r}) to $r \lesssim
10^{-8}$~\cite{Mizuno:2019bxy}; however, it remains out of reach for
upcoming experimental probes. Because Swampland program insights
suggest inflation requires extreme fine-tuning, it is natural to
explore alternative scenarios for the early universe. One such
possibility, rooted in string dualities and a topological phase for
the early universe, was proposed in~\cite{Brandenberger:1988aj} and
further explored in~\cite{Agrawal:2020xek}. Another possibility is the
ekpyrotic universe that undergoes repeating cycles of expansion and contraction~\cite{Khoury:2001wf,Khoury:2001bz,Steinhardt:2001st,Steinhardt:2002ih}.

Given the relation (\ref{HsqrtVrealtion}) between the Hubble parameter
and the potential, (\ref{TCCefolds}) implies  that $d$-dimensional scalar-field cosmologies in the asymptotic limit of field space must satisfy
\begin{equation}
  \left. \frac{|\nabla V|}{V}\right|_{\infty} \geq c_{\rm asym} \, ,
\label{TCCasym}
\end{equation}
where $c_{\rm asym} = 2/
\sqrt{d - 2}$ in reduced Planck units~\cite{Bedroya:2019snp}. Then, a remarkable consequence of the TCC is that if the
potential falls off exponentially, i.e.,
\begin{equation}
 \left. V(\phi) \right|_{_{{\rm lim} \, \phi \gg 1}} \sim \exp (-\gamma \, \phi) 
\end{equation}
then (\ref{TCCasym}) sets a restriction on the rate of fall off
$\gamma \gtrsim c_{\rm asym}$, which is exactly what is observed
asymptotically in field space in all the string landscape examples~\cite{Rudelius:2022gbz,Andriot:2022xjh,vandeHeisteeg:2023uxj}. The bound (\ref{TCCasym}) remains invariant under dimensional reduction~\cite{Rudelius:2021azq}. Furthermore, holographic arguments regarding infinite distance limits of field space independently motivate (\ref{TCCasym})~\cite{Bedroya:2022tbh}.  In
addition, if the universe undergoes a transition (which is driven by a
scalar field that rolls down a
potential) from an accelerating
expansion phase in the past to an eternal decelerating expansion phase
in the future, then the TCC must also be satisfied in the interior of moduli space~\cite{Bedroya:2024zta,Bedroya:2025ris}. Actually, for a monotonic potential within the range $[\phi_i, \phi_f]$, the TCC implies that $V$ is bounded above by an exponential function with a specific fall-off rate at every point of the interval~\cite{Bedroya:2019snp}
\begin{equation}
  V(\phi) \leq \frac{(d-1) (d-2)}{2} \exp \left(-\frac{2 \,  |\phi-\phi_i| }{\sqrt{(d-1) (d-2)}}\right) \, .
\label{TCCbulk}
\end{equation}
This can be shown using the $d$-dimensional Friedmann equation,
\begin{equation}
  \frac{(d-1)(d-2)}{2}H^{2}=\frac{1}{2}\ \dot \phi^{2}+V,
\end{equation}
along with the single-field equation of motion,
\begin{equation}
  \ddot \phi +(d-1)H \dot \phi +V'=0,
\end{equation}
where we are working in 
units where the reduced Planck mass is equal to one and $V'$ indicates the derivative of $V$ with respect to $\phi$. Given $V>0$, the Friedmann equation
implies the bound
\begin{equation}
  \frac{H}{|{\dot \phi}|}>\frac{1}{\sqrt{(d-1)(d-2)}} \, .
\end{equation}
Substituting this lower bound into the expression
\begin{equation}
\int _{\phi _{i}}^{\phi _{f}}\frac{H}{{\dot \phi }}d\phi =\int
_{t_{i}}^{t_{f}}Hdt<-\log H_{f},
\end{equation}
yields
\begin{equation}
  \frac{|\phi _{f}-\phi _{i}|}{\sqrt{(d-1)(d-2)}}<-\log H_{f} \,,
\end{equation}
which rearranges to the bound 
\begin{equation}
  H_{f}<\exp\left(-{\frac{|\phi _{f}-\phi _{i}|}{\sqrt{(d-1)(d-2)}}} \right).
\end{equation}
Finally, because the kinetic term is positive, this implies that the
potential is constrained by
\begin{equation}
  V< \frac{(d-1)(d-2)}{2} H^{2} \leq
\frac{(d-1) (d-2)}{2} \exp \left(-\frac{2 \,  |\phi-\phi_i|
  }{\sqrt{(d-1) (d-2)}}\right) \, .
\end{equation}
While (\ref{TCCbulk}) provides a slightly weaker constraint than (\ref{TCCasym}), allowing for a flatter potential, it holds true even inside the moduli space. The inequality  (\ref{TCCbulk}) suggests that even if the local gradient is small at certain points in the interior, it must be large enough on average across the trajectory 
\begin{equation}
M_p  \overline{\left|\frac{\nabla V}{V} \right|} \geq \gamma' = \frac{2}{\sqrt{(d-1) (d-2)}} \, 
\label{TCCbulk2}
\end{equation}
to satisfy TCC. It is important to note that $\gamma' \simeq 0.8$ when $d=4$.

\subsection{Constraints at Asymptotic Limits via the Emergent String Conjecture}
\label{sec:3d}

Numerous examples within string theory strongly suggest that any infinite distance limit corresponds to either a decompactification scenario or a tensionless string limit,
a proposal known as the emergent string conjecture
(ESC)~\cite{Lee:2019wij}.\footnote{For a teaching-focused discussion,
  see again~\cite{vanBeest:2021lhn}.} At a given infinite-distance limit in $d$ dimensions, the ESC sets the stage for bounding  $|\nabla \log m_{\rm lightest}|$~\cite{Etheredge:2022opl}
\begin{equation}
  \left| \frac{\nabla m_{\rm lightest}}{m_{\rm lightest}}\right| \geq
  \frac{1}{\sqrt{d-2}} \,
\label{asymmlightest}
\end{equation}
$|\nabla \log \Lambda_s|$~\cite{vandeHeisteeg:2023dlw,Bedroya:2025ltj} 
\begin{equation}
  \frac{1}{\sqrt{(d-1) (d-2)}} \leq \left|\frac{\nabla \Lambda_s}{\Lambda_s}\right| \leq \frac{1}{\sqrt{d-2}} \, .
\end{equation}        
and $|\nabla \log V|$~\cite{Bedroya:2025fie} 
\begin{equation}
  \left|\frac{\nabla V}{V} \right| \leq 2 \ \sqrt{\frac{d-1}{d-2}} \ .
\end{equation}

Castellano, Ruiz, and Valenzuela (CRV) demonstrated that the DC and the associated ESC dictate a generic pattern in the asymptotic behavior of EFTs that are consistent with quantum gravity~\cite{Castellano:2023stg,Castellano:2023jjt}. Specifically, in
infinite-distance limits the scalar product between the
(logarithmic) gradients of the mass gap of the lightest tower $m_{\rm lightest}$ and the species
scale $\Lambda_s$ satisfies a universal scaling relation
\begin{equation}
\frac{\nabla m_{\rm lightest}}{m_{\rm lightest}} \cdot \frac{\nabla \Lambda_s}{\Lambda_s} =
\frac{1}{d-2} \,,
\label{CRV}
\end{equation}
where $d$ is the number of dimensions of the lower dimensional
EFT. In terms of the number of species, using (\ref{species}) the CRV pattern can be recast as
\begin{equation}
  \frac{\nabla m_{\rm lightest}}{m_{\rm lightest}} \cdot \frac{\nabla N_s}{N_s} = -1 \,.
  \label{CRVpattern}
\end{equation}
Note that this relation is independent from the number of space-time dimensions.

An interesting connection of the CRV pattern and the quantum mechanics of the 1D theory obtained after dimensional reduction has been put forward elsewhere~\cite{Anchordoqui:2025izb}. Consider a dimensional reduction to one dimension ignoring metric fluctuations (i.e. assuming that the metric of the non-compact 4D space is Minkowski). Subject to these assumptions, the action (\ref{eft}) reduces to
\begin{equation}
  S_{\rm EFT} \supset \int dx^0 \left(\frac{1}{2} G_{ij} \dot{\phi}^i \dot \phi^j - V(\phi) \right)\, ,
\end{equation}    
where dots refer to $x^0$ differentiation. This action describes particles with field-space positions and proper time $x^0$. With this action, one can canonically quantize the position variables $\phi^i$ and their associated momenta $\pi_i$. Concretely, quantizing the scalars in one dimension, i.e., applying
the rules of quantum mechanics, we obtain
\begin{equation}
\pi_{\phi,i}={\delta {\cal L}_{\rm EFT}\over \delta(\partial_{x^0} \phi_i)}
=G_{ij} \dot \phi^j, \quad \partial_{x^0} \phi^j=G_{ij}  \dot \phi^j\, . \label{momentum}
\end{equation}
Taking this into account, the pair of conjugate variables $(\phi^i,\pi_{\phi,j})$  is subject to the following canonical equal-time commutation relation:
\begin{equation}
\lbrack \phi^i,\pi_{\phi,j}\rbrack =\lbrack \phi^i,G_{jk}\dot \phi^k\rbrack =i  \delta^i_j\, .
\end{equation}
As observed in~\cite{Anchordoqui:2025izb}, dot product relations between moduli-gradients can re-expressed in terms of commutation relations, e.g. in the context of the CRV pattern,  (\ref{CRVpattern}) corresponds to the imaginary part of a commutator.
\begin{equation}
\left\lbrack \log N_s, {d\over dx^0}\log m_{\rm lightest} \right\rbrack=i~
 {\nabla N_s\over N_s}\cdot {\nabla m_{\rm lightest} \over m_{\rm lightest}}\, . \label{fhcom1}
\end{equation}
As a consequence, if the CRV pattern holds, i.e.,
\begin{eqnarray}
\left\lbrack \log N_s, {d\over dx^0}\log m_{\rm lightest} \right\rbrack=
 -i\, , \label{fhcom2}
\end{eqnarray}
$N_s$ and the time derivative of $m_{\rm lightest}$ are related to canonically conjugate operators.

Because $|\Lambda| = V^2$, the AdS-DC (\ref{AdS-DC}) implies $|V| = m^{\mathtt{a}/2}$, and therefore $\nabla \log|V| = (\mathtt{a} \nabla \log m)/2$
which, via (\ref{fhcom2}), leads to
\begin{equation}
\left[\log N_s , \frac{d}{dx^0} \log \sqrt{V} \right]  = - \frac{2i}{\mathtt{a}} \,,
\end{equation}
with $2 < \mathtt{a} <4$. So this commutator 
 reaches its maximum possible extent through saturation of
the Higuchi bound, where $m_{\rm lightest}^2=V$. If this were the case, then 
$N_s$ and the time derivative of $\sqrt{V}$ would be related to canonically
conjugate operators. Note that a non-vanishing commutator requires
that the potential is not a constant, but varies with respect to $\phi$ inducing in this way a $x^0$-dependence. 

\subsection{Assessing Axion Fields as Drivers of Accelerated Expansion}

The axion field, $\phi$, originates as the angular component of a complex scalar field,
\begin{equation}
  \Phi = \zeta \ e^{i\phi/f_a}
\end{equation}
charged under a $U(1)$ symmetry. When the global $U(1)$ symmetry gets
broken, at the energy scale $f_{a}$, the radial field $\zeta$ acquires
a vacuum expectation value, $\langle \zeta \rangle = f_a/\sqrt{2}$,
and becomes heavy. As the Goldstone boson of this broken symmetry, the
axion is protected by a shift symmetry \mbox{$\phi \to \phi +
  \text{constant}$,} keeping it massless within perturbation
theory. However, non-perturbative (instanton) effects eventually break
this symmetry, generating an axion potential $V(\phi)$. This potential
must still respect a residual discrete shift symmetry, $\phi \to \phi
+ 2\pi n f_a$ for integer $n$, due to the axion's origin as the angle
of a complex field. Consequently, the resulting potential is
periodic. For an in-depth discussion of axion physics and their cosmological implications, see~\cite{Reece:2023czb} and~\cite{Cicoli:2023opf}, respectively.

From a string-theory perspective, axions are leading contenders for driving dark-energy due to their ubiquity and capacity to resolve inherent model-building challenges. Their suitability for dark energy stems from three key properties:
\begin{enumerate}[noitemsep,topsep=0pt]
  \item {\it Radiative stability:} Axion shift symmetries protect their potentials from being disrupted by quantum corrections.
\item {\it Scale Hierarchy:} Because their potentials are generated non-perturbatively, axions can operate at the tiny energy scales required for dark energy, even while other fields remain stabilized at much higher masses.
\item {\it Relaxed Constraints:} Their derivative couplings to other fields allow them to bypass strict experimental mass limits that typically constrain quintessence models.
\end{enumerate}  
To set the stage for axion-specific conjectures, we first need to cover the basics of the WGC, a major pillar of the entire swampland program.

The WGC requires that gravity acts as the weakest force in any consistent quantum gravity theory~\cite{Arkani-Hamed:2006emk,Harlow:2022ich}. This implies that for every $U(1)$ gauge field, there must exist a particle with a charge-to-mass ratio greater than or equal to that of an extremal black hole, i.e. $m \lesssim q M_p$. The WEC aims to prevent stable, non-supersymmetric black holes (extremal black holes) from violating thermodynamical laws by allowing them to decay via emitted light particles.

The {\it axion weak gravity conjecture} (AWGC) extends the WGC framework to pseudoscalar fields ~\cite{Rudelius:2015xta,Brown:2015iha,Bachlechner:2015qja,Heidenreich:2015wga,DiUbaldo:2026rly,Maldacena:2026jqd}.
 It posits that for axions with non-perturbative, instanton-induced
 symmetry breaking, the action of the axion-instanton pair must
 satisfy
\begin{equation}
  S_{\rm inst} \lesssim \frac{M_p}{f_a} \ .
\label{AWGC1}  
\end{equation}
For a theory to be described reliably by a non-perturbative potential
$V_{\rm eff}(\phi) \sim e^{-S_{\rm inst}} V(\phi)$, the instanton
action must satisfy $S_{\rm inst} \gtrsim 1$. Note that if the
instanton action were $S_{\rm
  inst} \ll 1$, the exponential suppression would fail, higher-order
instanton corrections would become large, and the EFT would lose predictive power.
This requirement together with (\ref{AWGC1}) places a necessary constraint on the theory, forcing the decay constant to be sub-Planckian, i.e. $f_a \lesssim M_p$.

 Very recently, Shiu, Tonioni, and Tran (STT) derived an analytic bound on the parameter space for axion dark energy by analyzing the phase space evolution of axion fields within a universe experiencing a decrease rate of cosmic acceleration~\cite{Shiu:2026edl}. The derivation links the axion mass and decay constant, showing the bound holds independently of initial misalignment angles.\footnote{The angular misalignment is a key, often random, initial value $\theta _{i}$ of the axion field relative to the minimum of its potential in the early universe, which determines the amount of dark matter produced.} In Sec.~\ref{sec:4b}  we demonstrate how the combined STT analytic bound and AWGC severely limit the role of axions in driving late-time accelerated expansion.

\subsection{The Bottom Line}

In short, within standard 4D cosmological frameworks, the Hubble
parameter acts as the measure of cosmic expansion. If an accelerated
expansion scenario is considered, then 
 (\ref{HsqrtVrealtion}) enables us to
set limits on the expansion rate based on the scalar
potential. Building on the derivation of these potential bounds using
swampland conjectures in Secs.~\ref{sec:3b}, \ref{sec:3c}, and
\ref{sec:3d}, the following sections provide a detailed examination of
these implications, while quintessence is confronted to DESI findings.

\section{Dynamics of Scalar Field Dark Energy}
\label{sec:4}

In this section we consider a cosmological model comprising a
minimally coupled, canonically normalized scalar field $\phi$, acting
as dark energy, alongside radiation and non-relativistic matter (cold
dark and baryonic). By selecting an appropriate field basis for this single-field scenario, the moduli space metric is reduced to a trivial form, enabling the relevant terms of the action (\ref{eft}) to be rewritten as
\begin{equation}
  S_{\rm EFT} =  \int d^4x \sqrt{-g} \ M_{p}^{2}\left(\frac{1}{2}
    {\cal R} - \frac{1} {2} {g^{\mu\nu} \ \partial_\mu \phi \,
      \partial_\nu \phi } - V(\phi) + \mathscr{L}_{m,r} \right)\, ,
\end{equation}
where $g_{\mu \nu}$ is the FLRW metric tensor, with line element 
given by (\ref{metric}). For a FLRW background, $\phi = \phi(t)$,
and thus the scalar field
Lagrangian density,
\begin{equation}
\mathscr{L}_{\phi }= - \frac{1}{2} \, g^{\mu \nu } \, \partial _{\mu }\phi \,
\partial _{\nu }\phi - V(\phi ) \,,
\label{Lphi}
\end{equation}
simplifies to
\begin{equation}
  \mathscr{L}_{\phi }=\frac{1}{2}\dot{\phi }^{2} - V(\phi ) \, .
\end{equation}  
The first Friedmann equation takes the form
\begin{equation}
  \left(\frac{\dot a}{a} \right)^2 = \frac{1}{3 M_p^2} \biggl(\rho_{m } +  \rho_{r} + \rho_\phi \biggr) - \frac{k}{a^2}
\end{equation}
and the second Friedman equation (\ref{Frie2}) becomes
\begin{equation}
  \frac{\ddot a}{a} = - \frac{1}{6 M_p^2} \ \biggl(\rho_{m } + 2 \rho_{r} + \rho_\phi (1 + 3 w_\phi) \biggr) \, .
\end{equation} 
The energy density $\rho_\phi$ and pressure $P_\phi$ of a canonical quintessence scalar field are derived by treating the field as a perfect fluid in a FLRW metric, leading to
\begin{equation}
  \rho_\phi = \frac{1}{2} \dot \phi + V(\phi)
\end{equation}
and
\begin{equation}
  P_\phi = \frac{1}{2} \dot \phi - V(\phi) \, .
\end{equation}
The field obeys the continuity equation,
\begin{equation}
  \dot \rho_\phi + 3 H (\rho_\phi + P_\phi) = 0,
\end{equation}
and its equation of motion
\begin{equation}
\ddot \phi + 3H \dot \phi + V' (\phi) = 0 \,,
\end{equation}
where for a single field, the operator $\nabla$ simplifies to a prime-denoted derivative  $\equiv \partial _{\phi}$.

The equation of state is given by
\begin{equation}
  w_\phi = \frac{P_\phi}{\rho_\phi} = \frac{\dot \phi/2 - V(\phi)}{\dot \phi/2 + V(\phi)} \, .
\label{wphi}
\end{equation}
It describes how the field acts as a fluid, with $w_\phi$ varying between $-1$ (potential-dominated) and $+1$ (kinetic-dominated).

Finally, the expression for the Hubble parameter is
\begin{equation}
  H^2 = \frac{1}{3M_p^2} \left(\rho_{m, {\rm today}} \ a^{-3} + \rho_{r,{\rm today}} \ a^{-4} + \frac{1}{2} \dot \phi^2 + V(\phi) \right) - \frac{k}{a^2} \,,
  \end{equation}
and the cosmic acceleration can be expressed as
  \begin{equation}
    \frac{\ddot a}{a} = - \frac{1}{6 M_p^2} \biggl(\rho_{m,{\rm today}} \ a^{-3} + 2 \rho_{r,{\rm today}} \ a^{-4} + 2 \dot \phi^2 - 2 V(\phi)  \biggr) \, .
\end{equation}
Moving on, we consider the influence of the distinct potentials.
  
\subsection{Exponential Quintessence: Bridging Steep Potentials, Curvature, and Strings}

\begin{table}
    \caption{The $1\sigma$ (i.e., $68.3\%$ confidence level) constraints on model parameters obtained through MCMC scans of the parameter space for the models and parametrizations studied in~\cite{Akrami:2025zlb}, along with their corresponding minimum $\chi^2$ and $\Delta\log B$, where $\Delta\log B_X = \log B_X- \log B_{\Lambda \mathrm{CDM}}$ for a given model or parametrization $X$. $\Delta\log B_X>0$ implies evidence in favor of model $X$ over $\Lambda$CDM, while $\Delta\log B_X<0$ implies evidence in favor of  $\Lambda$CDM over model $X$.
    These can be interpreted through the updated Jeffreys' scale:
    $\Delta\log B_{X}<1.1$ implies that model $X$ is comparable with
     $\Lambda$CDM, with neither one being distinctly preferred to
    the other; $1.1<\Delta \log B_{X}<3$ implies weak evidence favoring
    model $X$ to $\Lambda$CDM; $3<\Delta \log B_{X}<5$ implies
    moderate evidence favoring model $X$ to  $\Lambda$CDM; $\Delta
    \log B_{X,Y}>5$ implies strong support for model $X$ over 
    $\Lambda$CDM.  $N=1845$ data points were used in the statistical analysis. \label{tabla2}}
\renewcommand{\arraystretch}{1.3}
\rowcolors{1}{}{lightgray}
    \centering
    \setlength\tabcolsep{0pt}
    \begin{tabular}{ |c|c|c|c|c|c| }
    \hline
    \rowcolor{myblueLight}
    ~Parameter~ &  ~~ $\Lambda$CDM~~ & ~~$\Lambda$CDM$+\Omega_{k}$~~ & ~~ $\phi$CDM~~ & ~~$\phi$CDM$+\Omega_{k}$~~ & ~~~ CPL~~~  \\
    \hline
    $\Omega_{\rm{m}}$ & ~~$0.305\pm 0.003$~~ & $0.306\pm 0.003$ & ~~$0.315\pm 0.005$~~ & $0.316\pm 0.006$ & $0.320\pm 0.006$ \\
    $\Omega_{k}$ & $\cdots$ &$0.003\pm 0.001$ & $\cdots$ & $0.003\pm 0.001$ & $\cdots$\\
    $H_0$& $67.96\pm 0.23$ & $68.48\pm 0.30$ & $66.81\pm 0.56$ & $67.29\pm 0.62$ &$66.73\pm 0.57$\\
    $\kappa$ &$\cdots$&$\cdots$ & $0.698^{+0.173}_{-0.202}$  & $0.722^{+0.182}_{-0.208}$ &$\cdots$\\
    $V_0$ &$\cdots$&$\cdots$ & $2.207\pm 0.389$  & $2.299\pm 0.332$ & $\cdots$\\
    $w_0$ & $-1$ & $-1$& $\cdots$& $\cdots$& ~ $-0.751\pm 0.058$\,\\
    $w_a$ & 0 & 0& $\cdots$& $\cdots$& ~$-0.877\pm 0.231$\\
    $\chi^2$ & $1680.70$ & $1672.08$ & $1673.98$ & $1664.11$ & 1660.65\\
    $\Delta\log B$ & 0 & $-1.55$ & 4.03 & 3.55 & 6.84 \\
    \hline
    \end{tabular}
\end{table}

Cosmological scenarios featuring a quintessence field with an
exponential potential, $V(\phi) = V_0 e^{-\kappa \phi}$, show a slight
to moderate advantage in fitting current observational data over the
standard $\Lambda$CDM model~\cite{Ramadan:2024kmn}. In addition, it has long been known that the presence of negative spatial curvature $k=-1$ in a FLRW metric can significantly change the phase space dynamics, acting as a {\it brake} on the scalar field, which allows it to mimic a flatter potential and generate an epoch of acceleration~\cite{vandenHoogen:1999qq,Gosenca:2015qha}. Indeed, negative curvature allows the model to reach a {\it fixed point} in phase space that does not exist in a flat $k=0$ universe, leading to a {\it curvature-assisted} acceleration, particularly when the field is rolling down a steep potential. This mechanism is closely related to supergravity and string theory, as it attempts to reconcile the need for a steep potential (often found in string theory compactifications) with the required slow-roll behavior for dark energy~\cite{Marconnet:2022fmx,Andriot:2023wvg,Andriot:2024jsh}.

The stringy curvature-assisted model was tested against combinations
of DESI DR1 + CMB + SN Ia~\cite{Alestas:2024gxe,Bhattacharya:2024hep}
and DESI DR2 + CMB + SN Ia~\cite{Akrami:2025zlb,Dinda:2025iaq}, with
all analyses yielding highly consistent results. Table~\ref{tabla2}
displays the results from the MCMC likelihood analysis under DESI DR2
+ CMB + DESY5 in~\cite{Akrami:2025zlb}.
While exponential quintessence models generally show moderate
preference over $\Lambda$CDM, the dataset combination from
Table~\ref{tabla2} yields stronger results. Specifically, the
$\phi$CDM model is favored with $\Delta \log B \sim 4$ (about
$3.3\sigma$ significance), and the $\phi$CDM+$\Omega _{k}$ model is
favored with $\Delta \log B \sim 3.5$ (about $3.2\sigma$
significance). This statistical preference remains largely unchanged
even when incorporating two scalar fields with double-exponential
potentials~\cite{Alestas:2025syk}.

We conclude by contrasting these outcomes against the swampland
conjectures. Comparing the results in Table~\ref{tabla2} with (\ref{TCCasym}) and
(\ref{TCCbulk2}) setting $d=4$ reveals that although these two
quintessence models violate the asymptotic TCC bound, they remain
consistent with the TCC bound in the interior of moduli space within
$1\sigma$.

\subsection{The Quintessential Axion}

\label{sec:4b}

Axions offer a natural, string-theory-motivated mechanism to drive late-time acceleration~\cite{Choi:1999xn} or, if active earlier, potentially resolve the $H_0$ tension by increasing the expansion rate before recombination~\cite{Poulin:2018cxd}.
The potential for a single axion field $\phi$ is explicitly given by,  
\begin{equation}
  V (\phi) = \Lambda_a^4 \ [ 1 + \cos(\phi/f_a)]^n \,,
\label{Vaxion}
\end{equation}
with $n=1$ and where $\Lambda_a$ is the non-perturbative scale that is related to the axion mass by $\Lambda_a^2 \approx m_a f_a$~\cite{Freese:1990rb,Frieman:1995pm}. Potentials of the form in Eq.~(\ref{Vaxion}) are ubiquitous in string compactifications~\cite{Svrcek:2006yi,Arvanitaki:2009fg,Cicoli:2012sz,Marsh:2015xka}.

\begin{table}
  \caption{List of physical parameters and their constraints at 95\% CL using DESI DR1 + CMB +  different datasets of SN Ia. The columns are listed in terms of the SN Ia data sample~\cite{DESI:2025hce}. 
\label{tabla3}}
\renewcommand{\arraystretch}{1.3}
\rowcolors{1}{}{lightgray}
    \centering
    \setlength\tabcolsep{0pt}
    \begin{tabular}{ |c|c|c|c| }
    \hline
      \rowcolor{myblueLight}
      Parameters & ~~~~+ PantheonPlus~~~~ & ~~~~+ Union3~~~~ & ~~~~+DESY5~~~~ \\
\hline
      $w_{\phi,{\rm today}}$ & $-0.92^{+0.06}_{-0.05}$ & $-0.75^{+0.22}_{-0.18}$  & $-0.84^{+0.10}_{-0.08}$ \\
   $\Omega_{m,{\rm today}}$ & $0.314^{+0.011}_{-0.010}$ & $0.330^{+0.018}_{-0.016}$ & $0.322^{+0.014}_{-0.013}$ \\
 ~~~~$H_0/(10^{-33}~{\rm eV})$~~~~ & $1.43^{+0.02}_{-0.02}$ & $1.39^{+0.03}_{-0.04}$& $1.41^{+0.02}_{-0.03}$ \\
   $\log_{10}(m_a/{\rm eV})$ & $-32.69^{+0.18}_{-0.18}$ & ~~~~~~~~$ -32.48^{+0.20}_{-0.24}$~~~~~~~~ & ~~~~~~~~$-32.58^{+0.20}_{-0.21}$~~~~~~~~\\
$\log_{10}(f_a/M_p)$ & $-0.11^{+0.24}_{-0.22}$ & $-0.33^{+0.31}_{-0.23}$ & $-0.22^{+0.31}_{-0.26}$ \\
      \hline
    \end{tabular}
 
  \end{table}

Using DESI
DR1 results, CMB observations, and the three SN Ia datasets the DESI
Collaboration reported  constraints on the free parameters of the model~\cite{DESI:2025hce}. The results are encapsulated in Table~\ref{tabla3}. The constraints demand that the field starts in the hilltop regime, with initial condition 
\begin{equation}
\theta_i = \phi_i/f_a \sim 0.7 \, .
\label{phii}
\end{equation}
Note that (\ref{phii})  results from a multi-parameter likelihood analysis using the potential  (\ref{Vaxion}), with a prior on initial conditions of $\phi_i/f_a > 0.01$~\cite{DESI:2025hce}.

\begin{figure}[tpb]
\postscript{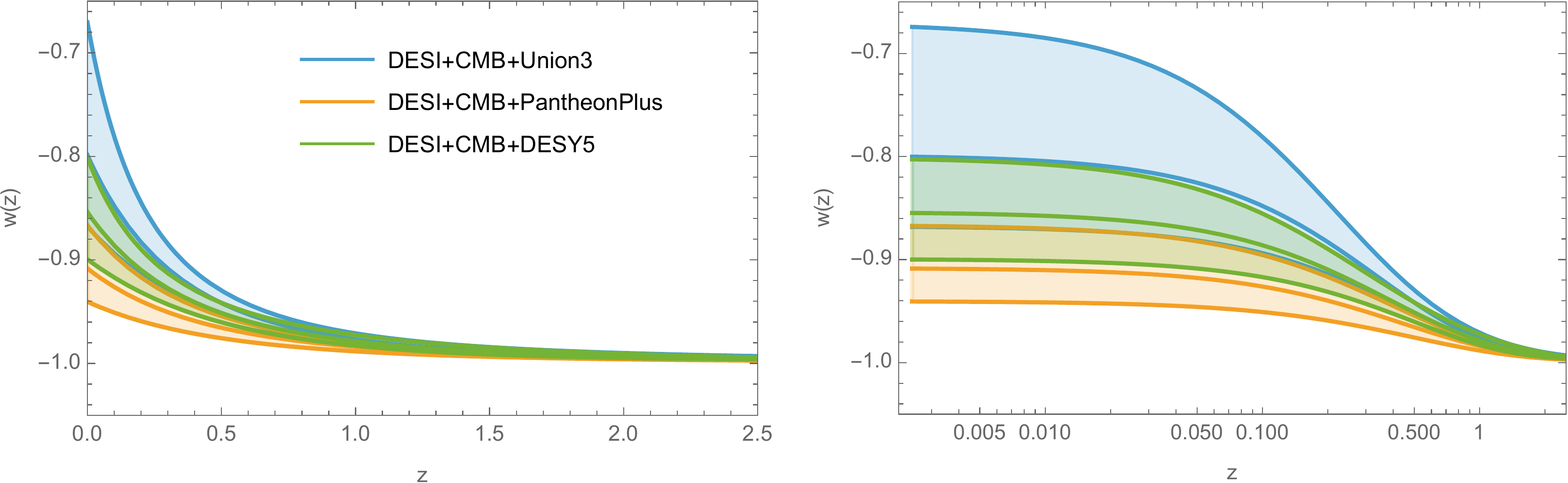}{0.95}
\caption{Marginalized constraints on the equation of state parameter $w(z)$ in linear (left) and logarithmic (right) scales. The contours on the right have been taken from Fig.~12 in~\cite{DESI:2025fii}.  \label{fig:1}}
\end{figure}

The integration of DESI DR2, CMB, and SN Ia datasets produces consistent results~\cite{DESI:2025fii}. Figure~\ref{fig:1} exhibits the marginalized posterior
distribution for the equation of state parameter associated with the scalar field potential (\ref{Vaxion}),
obtained using DESI DR2, CMB, and the three SN IA compilations.
The field rolls down the potential, reaching the present value of
$\phi_{\rm today}/f_a \sim 1.1$, traversing approximately $\Delta \phi
\sim 0.4 M_p$.

To summarize, simple quintessence models employing hilltop potentials,
which introduce two extra parameters beyond $\Lambda$CDM, generally align well with combined DESI DR2, CMB, and SN Ia data~\cite{DESI:2025fii}. While the +Pantheon and +Union3 datasets
provide only marginal evidence for these models over standard
$\Lambda$CDM, the +DESY5 compilation favors quintessence over
$\Lambda$CDM with approximately
$2.7\sigma$~\cite{Cline:2025sbt}.\footnote{Explicitly, for two degrees of freedom,
  the $\Delta \chi^2$ distribution simplifies to an exponential distribution
  $f(x;2) = e^{-x/2}/2$ and thus the $p$-value is given by $p = \int_{\Delta \chi^2}^\infty
  f(x;2) dx = e^{\Delta \chi^2/2}$. For $\Delta \chi^2 =
  10$~\cite{Cline:2025sbt}, we have
  $p \sim 0.0068$, which represents roughly a $2.7
  \sigma$ significance level; see Appendix~B
  in~\cite{Anchordoqui:2018qom}. This matches the estimate
  in~\cite{Bayat:2025xfr}.} There is slight ($\sim 2\sigma$) preference for
quintessence over $\Lambda$CDM when considering DESI DR2+CMB+DES-Dovekie~\cite{Shlivko:2025krk}.

Very recently, the results reported in~\cite{DESI:2025hce} were confronted with the swampland constraints~\cite{Shiu:2026edl}. The outcome of this analysis suggests a tension between current data and theoretical predictions for the axion mass. Specifically, the intersection of sub-Planckian decay constants (via AWGC) and the STT universal lower bound on the time from matter-dark energy equality to today yields an axion mass $m_a \sim {\cal O} (100) H_0$ that is roughly 100 times larger than current data suggest, see Table~\ref{tabla3}. This creates a severe challenge for embedding phenomenological axion dark energy models within UV-complete frameworks.

We end with two observations: {\it (i)}~
two-field early dark energy featuring $n=3$ axion-like potential
(\ref{Vaxion}) reduces the $H_{0}$ tension to a $1.5\sigma$
residual~\cite{Bella:2026zuk}; {\it (ii)}~preference for phantom crossing in DESI data can be realized within a kinetically mixed axion–dilaton quintessence model, a string-motivated system in which an axion-like field couples exponentially to a dilaton-like (moduli) field~\cite{Toomey:2025yuy}.

\subsection{$S$-Duality in Quintessence Models}

Gauge theory dualities are remarkable for connecting strongly coupled, non-perturbative field theories to weakly coupled, manageable descriptions. These relationships allow us to analyze strongly interacting systems by mapping them to weakly coupled duals. Consequently, dualities imply that two different classical limits can describe a single underlying quantum system. The $U(1)$ gauge theory on $\mathbb{R}^{4}$ is a textbook example, where electric-magnetic duality transforms the coupling constant and extends to an $SL(2, \mathbb{Z})$ action~\cite{Montonen:1977sn}. $S$-duality in string theory has been investigated in several contexts, see e.g.~\cite{Ferrara:1989bc,Font:1990gx,Cvetic:1991qm,Sen:1994fa,Alvarez-Gaume:1996ohl,Gopakumar:2000na,Nekrasov:2004js,Argyres:2007cn,Gaiotto:2008ak,Dimofte:2011jd,Leedom:2022zdm,Casas:2024jbw,Kallosh:2024ymt,Kallosh:2024pat,Cribiori:2024qsv}. In this subsection we review the compatibility of $S$-dual quintessence models with swampland conjectures and DESI findings.  We emphasize that we do not imply a direct connection to a specific string vacuum; rather, we view the self-dual constraint as a potential relic of string physics governing the late-time acceleration of the Universe.

For a real scalar field $\phi$, the $S$-duality symmetry
takes the form $\phi \to
-\phi$  (or analogously \mbox{$g\rightarrow 1/g$}, with $g\sim
e^{\phi/M_p}$). The $S$-duality constraint forces a
particular functional form on the potential:
$f[\cosh(\phi/M_p)]$~\cite{Anchordoqui:2014uua,Anchordoqui:2021eox}. The simplest $S$ self-dual form for the potential of the
quintessence field is given by
\begin{equation}
V(\phi) = \Lambda \ {\rm sech}
(\varkappa \, \phi/M_p) \,,
\label{Vsdual}
\end{equation}
where $\varkappa$ is an order one parameter. Actually, it is natural to take $\varkappa =
\sqrt{2}$, and therefore it is not an extra parameter of the potential as
it saturates the asymptotic TCC bound.

By comparing the Maclaurin series of $f(x) = {\rm sech}(\sqrt{2}x)$ and
$g(x) = [1 + \cos(2x)]/2$ around $x=0$, it is clear that they share
the same first two terms ($1 - x^2$). This makes $f(x)$ an accurate
quadratic approximation of $g(x)$ near the origin, with their leading
difference being $f(x) - g(x) \approx x^4/2$. Thus, for $f_a/M_p \sim 1/2$, the $S$-dual potential acts as an accurate
approximation of the axion potential.
Remarkably, because the value $f_a/M_p = 1/2$ lies within the
$1\sigma$ confidence region of the DESI Collaboration's
findings~\cite{DESI:2025fii}, setting the vacuum energy to $\Lambda \sim 10^{-120} M_p^4$ makes the
the potential (\ref{Vsdual}) virtually identical to the axion potential
(\ref{Vaxion})~\cite{Anchordoqui:2025fgz}. Consequently, $S$-dual quintessence yields a similar level of rejection significance against $\Lambda$CDM when evaluated with combined DESI DR2 + CMB + SN Ia data.

In addition,
\begin{equation}
\frac{|V'(\phi)|}{V(\phi)} = \sqrt{2} \tanh (\sqrt{2} \phi/M_p) \, . 
\end{equation}
Thus, since $\phi_i / M_p \gtrsim 0.5$, the expression
$\sqrt{2} \tanh(\sqrt{2} \phi / M_p) \gtrsim \gamma'$ holds for the
entire trajectory, and within the bulk of the moduli space, the S-dual
potential also satisfies the TCC bound (\ref{TCCbulk2}). Using the substitution $s=e^{\sqrt2 \phi/M_p}$, the $S$-dual potential can be rewritten as (\ref{Vsdual})
\begin{equation}
V (s) =    \frac{{2 \ \Lambda}}{s+1/s} \, .
\label{Vsdual2}
\end{equation}
Around the $\mathbb{Z}_2$-symmetric (self-dual point) point $\phi=0$ ($s=1$), we have
\begin{equation}
    \frac{V'}{V}\bigg\vert_{\phi=0}=0 \qquad {\rm and} \qquad \frac{V''}{V}\bigg\vert_{\phi=0}=-\frac{2}{M_p^2} ,
\end{equation}
in agreement with the dSC.\footnote{For a general overview of hilltop potentials in the context of the swampland conjectures, see~\cite{Storm:2020gtv}.} Besides, the potential changes concavity at
\begin{equation}
  \phi^*/M_p = \frac{\sqrt 2}{2}\log(\pm 1 + \sqrt{2}) \simeq 0.62 \, ,
\end{equation}
and could act as a minimal three-hyper-surface phase-space geometry. This specific structural shape provides the exact physical architecture required to drive the sequential phases of frozen evolution, early dark energy, depletion, quiescence, and late-time reactivation~\cite{Giare:2026tyk}.

A genuine concern about $S$-dual quintessence is that placing the
scalar field near the potential's peak requires {\it unnatural}
fine-tuning. However, recent work suggests this starting point might be physically motivated if the peak coincides with an enhanced
symmetry point, such as the $\mathbb{Z}_{2}$ point in $S$-dual
models~\cite{Chen:2025rkb}. In this scenario, the initial state mimics
spontaneous symmetry breaking. At the high temperatures of the early
universe, symmetries are typically restored, naturally driving the
field toward the symmetric peak. As the universe cools, this symmetry
breaks, and the field begins its descent. While quantum fluctuations
alone are too small (of order $\phi_i \sim H_0/M_p \sim 10^{-60}$)  to
move the field significantly within the current age of the universe,
thermal fluctuations (or a non-inflationary misalignment mechanism~\cite{Brandenberger:2025axw})  likely provided the initial {\it push} needed to reach the observed late-time cosmological conditions. Note that it takes $t \sim 1/H_0 \
\log (M_p/H_0)$ for the
field $\phi$ to reach values of order
unity~\cite{Vafa:2025nst,Rudelius:2019cfh}, which is roughly 100 times more than the age of the universe.

 \subsection{Dynamics of Quintessence under Modular Symmetry}

 The main feature of the scalar potential (\ref{Vsdual}) is its symmetry under $\phi \to -\phi$, or more intuitively, \(s \to 1/s\). To expand this into a full modular group invariance, we first redefine $s$ as a complex scalar field $S$
 \begin{equation}  S = s + ia = e^{\tilde{\phi}} + ia, \quad
   \text{where} \quad \tilde{\phi} = \sqrt{2} \phi / M_p  \, .
 \end{equation}
 Under $SL(2, \mathbb{Z})$, this field transforms as
 \begin{equation}
   \label{eq:modtranf}
   S \to \frac{aS + ib}{icS + d} \, .
 \end{equation}
 To build a modular invariant function from \(S\), we use the Dedekind eta function $\eta(S)$. Specifically, the combination\begin{equation} \label{eq:modinfcomb}\left|\eta(S)\right|^4 (S + \bar{S})  \end{equation}
remains invariant under these transformations.
Given the asymptotic $\eta(S) \to e^{-\frac{\pi}{12}{\rm Re}S}$ and $|\eta(S)|^{4} (S+\bar S) \to e^{-\frac{n\pi}{3}{\rm Re} \, S}(S+\bar S)$ as ${\rm Re} \ S \to \infty$, an educated guess for the modular-invariant scalar potential is
\begin{equation} V(S, \bar S)  = -\frac{ \Lambda \ \log\left[|\eta(1)|^{4} \, 2\right]}{\log [|\eta(S)|^{4} \ (S+\bar  S)]} \, ,
\label{VmodularS}
\end{equation}
normalized such that $V(1, 1)= \Lambda$ using $\eta (1)=\Gamma
\left(1/4\right)/(2\pi ^{3/4})$.  In terms of $\phi$ the modular invariant potential can be recast as
\begin{equation}
  V(\phi) = \frac{\Lambda \log(|\eta(1)|^4 2)}{\log\left[|\eta (e^{\sqrt{2} \phi})|^4 (e^{\sqrt{2}\phi}+e^{-\sqrt{2} \phi})\right]} \, .
\label{Vmodular}
 \end{equation}      

\begin{figure}[htb!]
  \begin{minipage}[t]{0.49\textwidth}
    \postscript{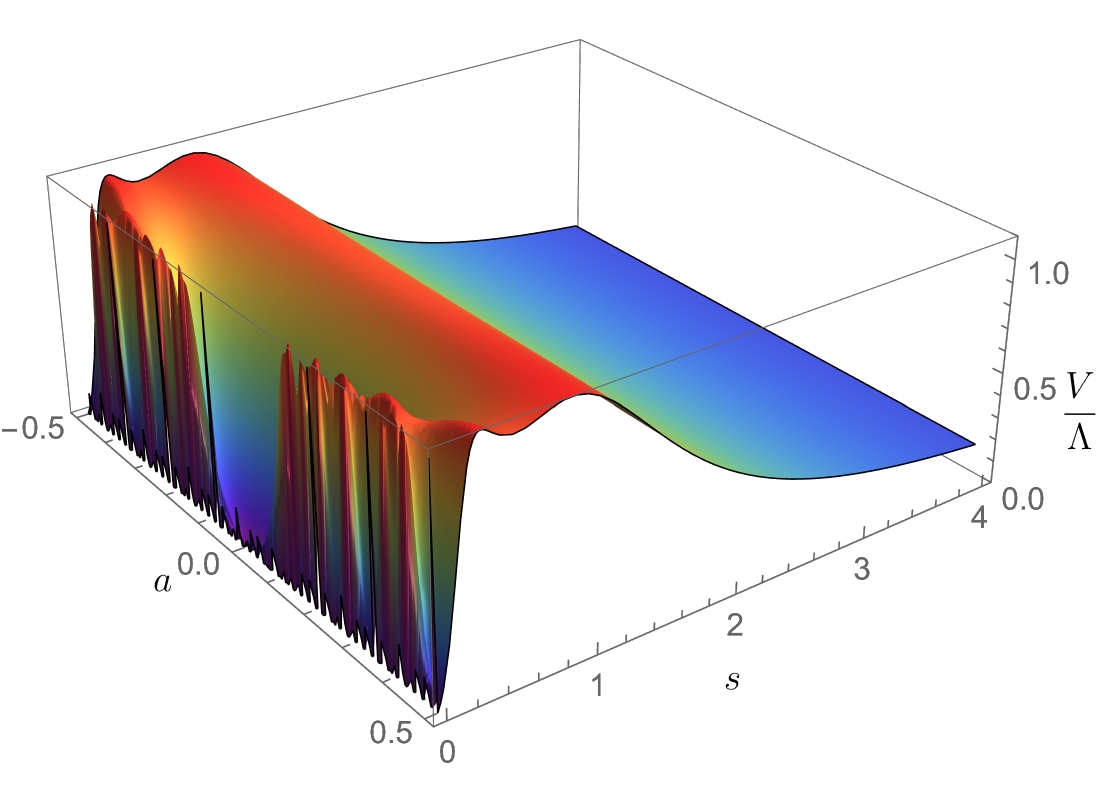}{0.99}
  \end{minipage}
\begin{minipage}[t]{0.49\textwidth}
    \postscript{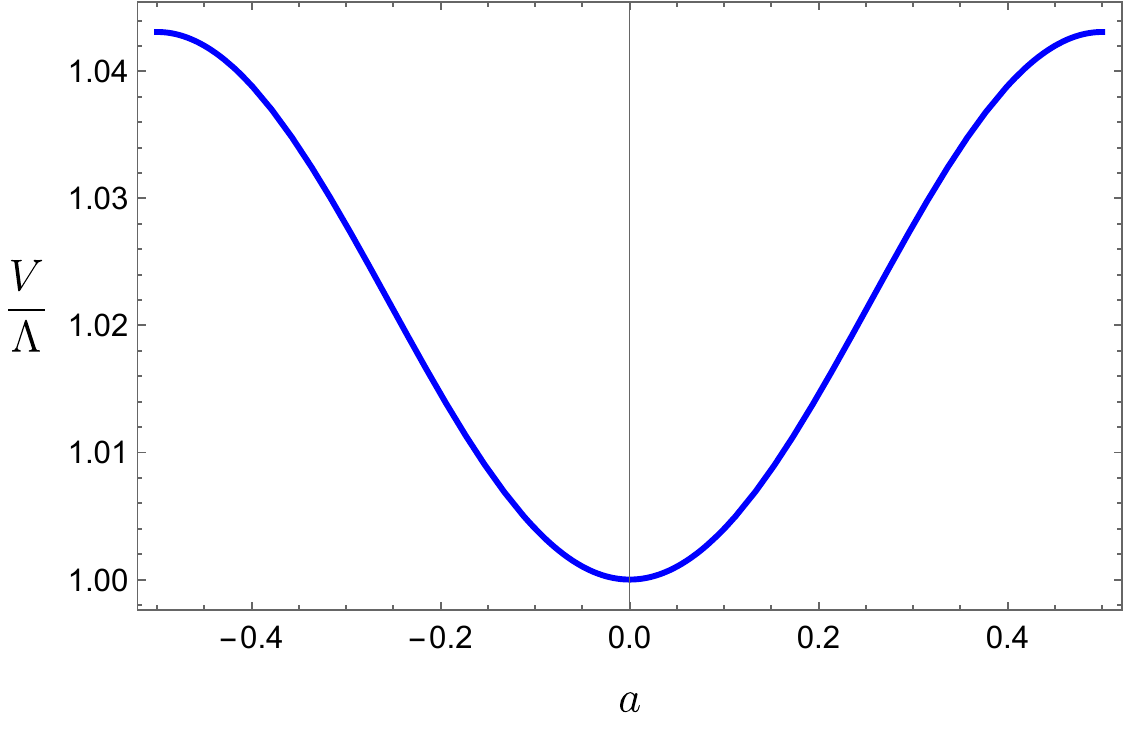}{0.99}
  \end{minipage}
\caption{Modular invariant potential (\ref{VmodularS}) in the
  $s$-$a$ plane (left) and along the axionic direction
  for $s=1$ (right). At $a=0$ and $s=1$ the potential has a minimum which
  persists for $s \geq 1$. \label{fig:2}}
\end{figure}

As shown in Fig.~\ref{fig:2}, the  profile of the $V(S,\bar S)$
potential in the $a$–$s$ plane
illustrates modular invariance, characterized by infinite copies of
the fundamental domain below $s < 1$. Oscillations exist along the
$a$-direction, but with only $\sim 4\%$ impact, they are difficult to
discern at $s = 1$ in the left panel of Fig.~\ref{fig:2}. Remarkably, $a = 0$ serves as a consistent truncation because the
potential reaches a minimum there. When looking at the
$s=1$ slice along the $a$-direction, which is displayed in the right panel of
Fig.~\ref{fig:2}, we see that the minimum is at the self-dual
point. With the minimum persisting at the self-dual point $\forall s
\geq 1$, the axion $a$ is effectively frozen at the potential's minimum.

\begin{figure}[tpb]
\postscript{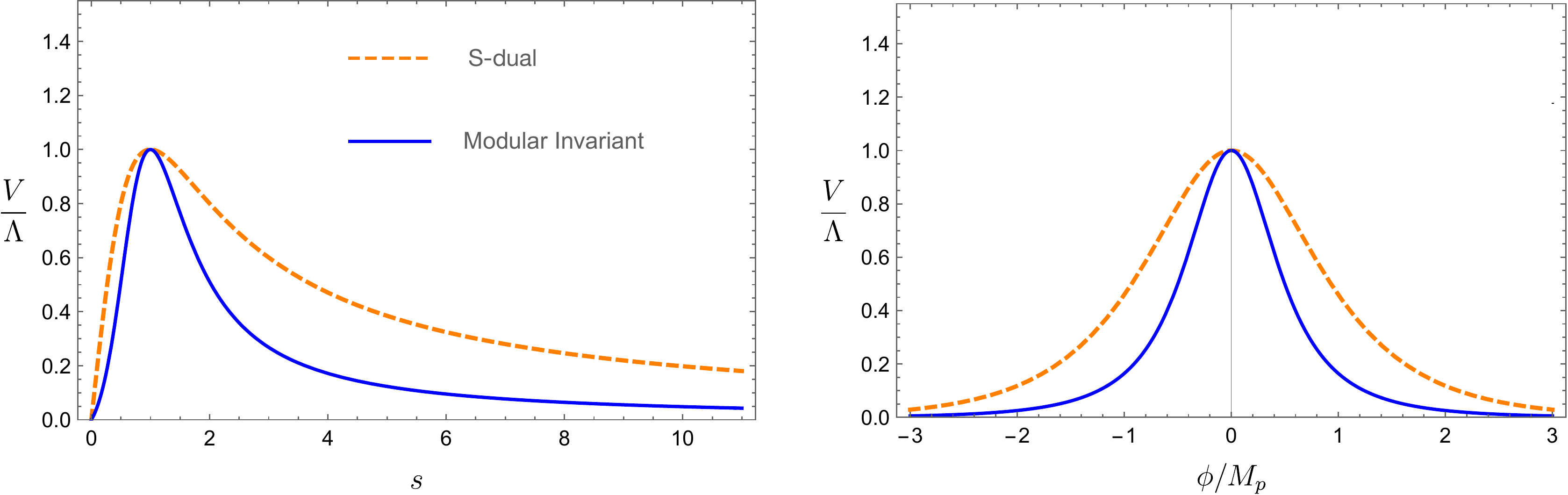}{0.95}
\caption{Comparison of the $S$-dual  (orange) and modular-invariant  (blue) potentials in the $s$ (left) and $\phi$ (right) coordinates.  \label{fig:3}}
\end{figure}

The modular invariant potential shares key characteristics with the $S$-dual potential (\ref{Vsdual}) at both the self-dual point and in the large-$s$ limit. The similarities between (\ref{Vsdual}) and (\ref{Vmodular}) are evident in Fig.~\ref{fig:3}.  It is worth noting that the modular invariant potential adheres to swampland constraints, as it stays below the $S$-dual potential. Modular invariant quintessence is consistent with DESI DR2 + CMB + Union3 at $1\sigma$, and it is just outside the $1\sigma$ confidence regions of DESI DR2  + CMB + PantheonPlus and DESI DR2 + CMB + DESY5~\cite{Anchordoqui:2025epz}.

We close with an intriguing observation connecting $S$-dual and
modular-invariant quintessence with the dark dimension scenario. These quintessence scenarios can be
described by a two-dimensional moduli space involving two scalar
fields, $R$ and $s$. In the simplest case, the scalar potential factorizes into two terms,
\begin{equation}
  V(R,s) = \Lambda(R) V(s) \,,
\end{equation}
where $R$ (located at the boundary)
determines the vacuum energy within the dark dimension scenario, and
$s$ (living in the bulk) acts as the quintessence field governing
the time variation of this energy. Indeed, currently the saxion $s$ is situated
near the self-dual point and rolls down towards an asymptotic value in
the far future.

\section{Non-Gravitational Dark Sector Coupling}

\label{sec:5}

The concept of a coupling between dark matter and dark energy dates back at least to the early 1990s~\cite{Casas:1991ky},
with interest in the subject spiking a decade
later~\cite{Amendola:1999er,Amendola:2000uh,Gasperini:2001pc,Comelli:2003cv,Farrar:2003uw,Hoffman:2003ru,Franca:2003zg,Majerotto:2004ji,Anchordoqui:2007sb}. Interacting
dark sector solutions for the Hubble tension gained
momentum~\cite{Salvatelli:2014zta,DiValentino:2017iww,Yang:2018euj,Kumar:2019wfs,Agrawal:2019dlm,DiValentino:2019ffd,
  Anchordoqui:2019amx,Lee:2022cyh}, but subsequent DESI DR1 results demonstrated
that these interactions naturally align within the favored $(w_0,
w_a)$ parameter
space~\cite{Giare:2024smz,Chakraborty:2024xas,Aboubrahim:2024cyk,Chakraborty:2025syu,Silva:2025hxw,Andriot:2025los,Bedroya:2025fwh,Pourtsidou:2025sdd,Li:2026xaz,
  Antusch:2026ldp,Gomez-Valent:2026ept,Wang:2026vqw}. In this section, we begin by discussing the phenomenology of dark matter-dark energy interactions, then turn our attention to the model detailed in~\cite{Bedroya:2025fwh}, which directly relates to the dark dimension and provides the best fit to the data.

\subsection{Phenomenology of Secluded Dark Sector Interactions} 

\label{sec:5a}

An interaction between dark matter ($dm$) and dark energy is usually
introduced by modifying the covariant
conservation of the energy-momentum tensor while keeping the Einstein equations intact. For a general model, the
energy transfer between $dm$ and $de$ is defined by the
exchange vector $Q^{\nu }$, with
\begin{equation}
 \nabla_\mu T^{\mu \nu}_{de} = - \nabla_\mu T^{\mu \nu}_{dm} = Q^\nu 
\end{equation}
that ensures the total energy-momentum tensor remains conserved,
$\nabla_\mu (T^{\mu \nu}_{de} + T^{\mu \nu}_{dm}) = 0$~\cite{Kodama:1984ziu}. Because the microscopic details of dark sector interactions are
unknown, the functional form of $Q^\nu$ is typically
phenomenological.

A widely studied framework is coupled quintessence (CQ), where the
interaction is defined as $Q^{\nu} = \beta 
T_{\mu,dm}^\mu \, \partial^{\nu}\phi $, with $\beta$ a dimensionless coupling constant that sets the magnitude and direction of energy transfer~\cite{Amendola:1999er}. For FLRW, this interaction reduces to
$Q = -\beta T_{\mu,dm}^\mu \, \dot \phi$. Alternatively, a simpler
phenomenological model is characterized by an interaction of the form $Q^\nu = Q
u^\nu_{dm}$, where $u_{dm}^{\nu}$ is the dm four-velocity, $Q = \xi H
\rho_{dm}$, and $\xi$ the coupling constant~\cite{Majerotto:2004ji}. For pressureless dm,
the resulting continuity equations for the dark sector components can be recast as
\begin{equation}
\dot{\rho }_{\rm CDM} + 3 H \rho_{\rm CDM}=- Q 
\label{dotrhoQ}
\end{equation}
and
\begin{equation}
  \dot{\rho }_{de}+3 H (1+w) \rho_{de}= Q \, ,
\end{equation}
with $Q = \beta \rho_{\rm CDM} \dot \phi$, or $Q = \xi H \rho_{\rm
  CDM}$. Note that if $Q > 0$ CDM releases
energy into dark energy, whereas if $Q < 0$ the energy flows in the
opposite direction.

Data from DESI, CMB, and SN Ia seem to favor a $Q \neq 0$ interaction
parameter~\cite{Giare:2024smz,Chakraborty:2024xas,Li:2026xaz}, though
one study offers a different
perspective~\cite{Gomez-Valent:2026ept}.\footnote{It is important to
  stress that the analysis of~\cite{Li:2026xaz} utilizes a combined
  dataset of {\it Planck}, ACT, and SPT, including joint lensing
  constraints, setting it apart from the study
  of~\cite{Gomez-Valent:2026ept} that relies solely on Planck
  data. This difference in data, particularly the high-multipole
  ($\ell > 4000$) sensitivity of ACT and SPT, accounts for
  improvements in the CQ model, as high-$\ell$ data could
  significantly impact interaction models~\cite{Brax:2023rrf}. In
  addition, despite the strong comparability of the underlying physics
  across the scalar-field models, their initial conditions feature
  different parameterizations. Namely, the method employed
  in~\cite{Gomez-Valent:2026ept} adopts an inverted power-law
  potential normalized to $\phi_{\rm initial}$, which is sampled as an
  extra parameter.} Particularly, for a potential
$V(\phi) \propto \phi^{-\alpha}$, the data strongly support a positive
coupling ($\beta > 0$) with a statistical significance exceeding
$5\sigma$ (see Table \ref{tabla4})~\cite{Li:2026xaz}. The preference
for positive $\alpha$ values points toward an inverse power-law
potential that decays as the field grows, resulting in a shallow slope
that facilitates a slowly evolving quintessence field. In this model,
dark matter dilutes faster than in \(\Lambda \)CDM due to the
interaction, while continuous energy transfer keeps the dark energy
density higher at intermediate redshifts
($0.5 \lesssim z \lesssim 2$). This shift causes the background
evolution to deviate from $\Lambda$CDM, effectively mimicking the CPL
$w$-parameter that appears {\it phantom-like} at earlier times before
transitioning back to a quintessence phase today. Importantly, this
phantom behavior is only an effective result of the interaction; the
underlying scalar field maintains a strictly quintessence-like
equation of state across all redshifts.

\begin{table}
  \caption{Comparison of the CPL and CQ models against $\Lambda$CDM
    across various combinations of DESI DR2 + CMB + SN Ia
    data~\cite{Li:2026xaz}. Goodness-of-fit is evaluated as $\Delta \chi^2_{\rm MAP}$,
    which measures the $\chi^{2}$ offset from $\Lambda$CDM at maximum
    a posteriori (MAP). For two degrees of freedom, $\Delta \chi^2_{\rm MAP} = -11.83 $ corresponds to a $3\sigma$ confidence level. Alongside are the $1\sigma$ constraints for
    the FDS $\alpha$ and $\beta$ parameters. \label{tabla4}}
\renewcommand{\arraystretch}{1.3}
\rowcolors{1}{}{lightgray}
    \centering
    \setlength\tabcolsep{0pt}
    \begin{tabular}{ |c|c|c|c|c| }
    \hline
      \rowcolor{myblueLight}
      Dataset Compilations & $\Delta \chi^{2^{\rm CPL}}_{\rm MAP}$ &
                                                                   $\Delta
                                                                   \chi^{2^{\rm
                                                                   CQ}}_{\rm
                                                                   MAP}$
     & $\alpha$ & $\beta$ \\
\hline
~~~~~DESI+CMB+PantheonPlus~~~~~ & $-9.88$ & ~~~$-9.11$~~~ & $0.41 \pm 0.19$  & $0.0517^{+0.0092}_{-0.0078}$ \\
      ~~~DESI+CMB+DESY5~~~ & ~~~$-20.74$~~~ & ~~~$-14.73$~~~ &
                                                                     ~~~$0.62
                                                               \pm
                                                               0.17$~~~
                & ~~~$0.0561^{+0.0094}_{-0.0068}$~~~ \\
      ~~~DESI+CMB+DES-Dovekie~~~ & ~~~$-14.13$~~~ &
                                                      ~~~$-13.66$~~~
                                                    
      &$ 0.46 \pm 0.18$ & ~~~$0.0527 \pm 0.0087$~~~ \\
\hline
    \end{tabular}
 
  \end{table}

  In contrast, the scenario in which $Q$ is proportional to the Hubble
expansion strongly prefers negative coupling ($\xi < 0$) and de phantom
equation of state ($w < -1$), triggering an energy transfer from de to dm~\cite{Li:2026xaz}. Consequently, de is suppressed while dm dilutes slower than in $\Lambda$CDM, yielding a high present-day matter abundance ($\Omega_m \sim 0.6$). To compensate for this at the background level, the phantom $w$-parameter allows the model to match distance measurements. At the perturbation level, this results in a heavily suppressed growth of structure. Despite lying in an extreme parameter space, the model’s best-fit solutions outperform $\Lambda$CDM and CPL against CMB, BAO, and SN data~\cite{Li:2026xaz}.

\subsection{Fading Dark Sector and the Dark Dimension}

As mentioned in Sec.~\ref{sec:3}, for a theory coupled to quantum gravity in dS space, the AdS-DC implies the existence of an infinite tower of states that becomes light as $|\Lambda| \to 0$~\cite{Lust:2019zwm}. To examine the AdS-DC via EFT, we consider a framework
comprising Einstein gravity with a cosmological constant, the SM, and additional fields. We focus on the infrared limit, investigating the dynamics as the dark energy density is taken to zero.
Schematically, in reduced Planck units, 
\begin{equation}
  S_{\rm EFT}  = \frac{1}{2}  \int d^4x \sqrt{-g} \ \left( {\cal R} - 2 \Lambda + \cdots ^{|\Lambda| \ll1}_{\xrightarrow{\hspace{0.5cm}}} \right) \, ,
\label{SEFTEL}
\end{equation}
where in the context of future objectives, hereafter $\Lambda >0$
represents a dark energy density that, rather than being strictly a
cosmological constant, may instead be a dynamic or rolling field.
While conventional EFT assumes a fixed setup, usually suggesting that
``nothing happens'' at low energies, string-theoretic UV consistency
forces us to include an infinite tower of new particles $\chi_n$ with masses
quantized according to the inverse compactification radius, $m_n \sim
n/R$, yielding
\begin{equation}
  S_{\rm EFT} =  \frac{1}{2} \int d^4x \sqrt{-g} \ \left({\cal R} - 2 \Lambda
   +  \sum_n \left(g^{\mu \nu} \partial_\mu \chi_n \, \partial_\nu
     \chi_n - m_n^2 \chi_n^2 \right) + \cdots \right) \, ,
\end{equation}
where the KK subscript has been omitted for brevity.
In this way, the AdS-DC underscores the significance of UV-IR mixing.

As mentioned in Sec.~\ref{sec:3a}, within the dark dimension scenario~\cite{Montero:2022prj} the KK graviton tower serves as a
strong candidate for dark matter\cite{Gonzalo:2022jac}. In particular, if the extra
dimension does not admit isometries the KK momentum of the graviton
tower is not conserved and hence a given KK mode of the tower could
decay into final states that include other, lighter KK excitations~\cite{Mohapatra:2003ah}. The cosmic evolution of the dark matter sector is
primarily driven by dark-to-dark decay
processes that govern the decay
of KK modes within the dark tower~\cite{Gonzalo:2022jac}. This implements a specific version
of the dynamical dark matter framework proposed
in~\cite{Dienes:2011ja}.

Focusing on the 4D EFT, the particle physics landscape consists of
three main components: {\it (i)}~the SM (originating from the brane),
{\it (ii)}~a tower of KK gravitons from the dark dimension, and {\it
  (iii)}~a set of scalar fields $\{\phi_i\}$ that controls the shape
and size of the compact space. In particular, the scale of the dark matter mass depends on the radion (the scalar
field controlling the length of the dark dimension). In other words,
if we assume the cosmological constant is dynamic, then the mass of
dark matter must also be dynamic, given its relationship to the
graviton tower's mass scale. Given these considerations, the dark dimension scenario offers a natural, unified framework that links dark matter to dark energy.

For phenomenological purposes, the dynamical dark energy component is modeled via a rolling scalar
field $\phi$, expressed as a linear combination of the $\{\phi_i\}$
components. A straightforward description of the interaction between dark matter
and dark energy involves local, exponential forms, where the potential
$V(\phi)$ and the cold dark matter mass $m_{\rm CDM}$ are defined
respectively as
\begin{equation}
  V(\phi) = V_0 \ e^{-c_1\phi}
\end{equation}
and 
\begin{equation}
  m_{\rm CDM} = m_0 \ e^{-c_2 \phi} , \end{equation}
and where the exponents $c_{1}$ and $c_{2}$ are of order one in
reduced Planck units, $m_0$ defines the initial dark matter mass scale, and $V_{0}$ dictates the scalar potential's energy scale~\cite{Bedroya:2025fwh}. These expressions are intended as local approximations for a field range smaller than $M_{p}$, rather than as universal, global functions.

The combined energy density and pressure for dark matter and dark energy are respectively expressed by
\begin{equation}
  \rho_{{\rm CDM+}de} = \frac{m_0 n_0}{ a^3} \ e^{-c_2\phi}+\frac{1}{2}\dot\phi^2+V_0e^{-c_1\phi}
\end{equation}
and
\begin{equation}
P_{{\rm CDM+}de} = \frac{1}{2}\dot\phi^2 - V_0e^{-c_1\phi}\,,
\end{equation}
where $n_0$ stands for the initial dark matter number density. The dark matter sector is defined by
\begin{equation}
  \rho_{\rm CDM} = \frac{\rho_{{\rm CDM},0}}{a^3} \ e^{-c_2\phi_i}
\end{equation}
and obviously $P_{\rm CDM} = 0$. The remaining components are attributed to dark energy
\begin{equation}
  \rho_{de}=\frac{1}{2} \dot{\phi }^{2}+V_{0}e^{-c_1\phi }+\frac{\rho _{{\rm CDM},0}}{a^{3}} \left[e^{-c_2 \phi }-e^{-c_2\phi_i}\right]
\end{equation}
and
\begin{equation}
  P_{de}=\frac{1}{2} \dot{\phi }^{2}-V_{0}e^{-c_1\phi } \, ,
\end{equation}
where $\phi _{i}$ represents the scalar field at the onset of dark matter mass evolution.

By defining CDM as a pressureless fluid scaling as $1/a^3$, the
field-dependent mass deviations are
attributed to an effective dark energy equation of state,
\begin{equation}
  w_{\rm eff} = w_\phi / (1 + x)
\end{equation}
with $w_{\phi }$ as given by (\ref{wphi}) and
\begin{equation}
x =  \frac{\rho _{{\rm CDM},0}}{a^{3}\rho _{\phi }}\left(e^{-c_2 \phi }-e^{-c_2 \phi _i}\right)
\end{equation}
In this formulation, if the scalar field increases ($\phi > \phi_i$), $x$ becomes negative, enabling $w_{\rm eff} < -1$ (phantom-like behavior) while keeping $w_\phi \geq -1$. The dynamics are governed by the Friedmann 
\begin{equation}
3M_p^2 H^2=\frac{1}{2}\dot\phi^2+V_{\rm{eff}}(\phi)+\frac{\Omega_r (3 M_p^2
  H_0^2)}{a^4} +\frac{\Omega_b (3 M_p^2 H_0^2)}{a^3}\,,\nonumber
\end{equation}
and field
\begin{equation}
\ddot\phi+3H\dot\phi+ V'_{\rm{eff}}=0\,,
\end{equation}
equations, with
\begin{equation}
V_{\rm{eff}}=V_0e^{-c_1\phi}+ \frac{\rho_{{\rm CDM},0}}{a^3} \ e^{-c_2\phi}\,,
\end{equation}
where $\Omega_r$ and $\Omega_b$ are the 
radiation and baryon density parameters, respectively.                                                

\begin{table}[ht!]
\caption{Statistical significance comparing the FDS model for $c_1 <
  0$ alongside the CPL parametrization, across various combinations of
  SN Ia datasets, to $\Lambda$CDM. Additionally, the mean, the $\pm
  1\sigma$ uncertainty, and the best fit value between parenthesis are
  provided for $c_1$ and $c_2$. 
\label{tabla5}}
\renewcommand{\arraystretch}{1.3}
\rowcolors{1}{}{lightgray}
    \centering
    \setlength\tabcolsep{0pt}
\begin{tabular}{|c|c|c|c|c|}
  \hline
\rowcolor{myblueLight}
Datasets  & ~~$\sigma^{\!\!\!\!\!\!\!^{\rm CPL}}$~~  & ~~$\sigma^{\!\!\!\!\!\!\!^{\rm FDS}}$~~ & $c_1$ & $c_2$\\
\hline
{+CMB} & $3.0$ &  {$2.5$}  & ~$-0.72^{+0.45}_{-0.43}$ ($-1.03$)~ & ~$0.05\pm0.01$ ($0.05$)~ \\ 
{+CMB+Union3} &  {$3.7$} & $3.4$ & ~$-1.01^{+0.21}_{-0.22}$ ($-1.12$)~ & ~$0.05\pm0.01$ ($0.05$)~ \\ 
~{+CMB+PantheonPlus}~  & {$2.7$} & {$2.9$} &  ~$-0.76^{+0.23}_{-0.22}$ ($-0.85$)~ & ~$0.05\pm0.01$ ($0.05$)~\\ 
{+CMB+DESY5}  & {$4.1$}  & {$4.0$} & ~$-1.00^{+0.13}_{-0.14}$
                                     ($-1.06$)~ & ~$0.05\pm 0.01$
                                                  (0.05)~\\
  ~{+CMB+Union3.1}~ &3.4 & 3.0 & $-0.87\pm 0.24$ ($-0.98$) & $0.05 \pm 0.01$
                                                            (0.05) \\
~~+CMB+PantheonPlus (recal)~~ & ~~~~3.2~~~~ & ~~~~3.2~~~~ & $-0.87 \pm 0.18$ ($-0.92$) &
                                                                      $0.05
                                                                      \pm
                                                                      0.01$ (0.05) \\
{+CMB+DES-Dovekie} & {$3.1$} & {$3.2$} & ~$-0.81 \pm 0.19$ ($-0.92$)~ & ~$0.05\pm 0.01$ (0.05)~\\ 
  \hline
\end{tabular}
\end{table}

\begin{table}[ht!]
\caption{Statistical significance comparing the FDS model for $c_1 >
  0$ alongside the CPL parametrization, across various combinations of
  SN Ia datasets, to $\Lambda$CDM. Additionally, the mean, the $\pm
  1\sigma$ uncertainty, and the best fit value between parenthesis are
  provided for $c_1$ and $c_2$. \label{tabla6}}
\renewcommand{\arraystretch}{1.3}
\rowcolors{1}{}{lightgray}
    \centering
    \setlength\tabcolsep{0pt}
\begin{tabular}{|c|c|c|c|c|}
  \hline
\rowcolor{myblueLight}
Datasets &  ~~$\sigma^{\!\!\!\!\!\!\!^{\rm CPL}}$~~ &  ~~$\sigma^{\!\!\!\!\!\!\!^{\rm FDS}}$~~ & $c_1$ & $c_2$\\
\hline
{+CMB}  & $3.0$  & {$2.1$}  & ~$0.43^{+0.28}_{-0.29}$ ($0.42$)~ & ~$0.05\pm0.01$ ($0.05$)~ \\ 
{+CMB+Union3} &  {$3.7$}  & $2.9$ & ~$0.71^{+0.22}_{-0.24}$ ($0.81$)~ & ~$0.05\pm0.01$ ($0.06$)~ \\ 
~{+CMB+PantheonPlus}~ &  {$2.7$}  & {$2.6$} &  ~$0.54^{+0.21}_{-0.24}$ ($0.65$)~ & ~$0.05\pm0.01$ ($0.05$)~\\ 
{+CMB+DESY5} &  {$4.1$} & {$3.6$} & ~$0.79^{+0.14}_{-0.12}$ (0.83)~ &
                                                                      ~$0.05\pm 0.01$ (0.06)~\\
  ~{+CMB+Union3.1} & 3.4 & 2.5 & $0.60 \pm 0.24$ (0.72) & $0.05 \pm
                                                         0.01$ (0.05)
  \\~~~{+CMB+PantheonPlus (recal)}~~~ & ~~~~3.2~~~~ & ~~~~2.9~~~~ & ~~$0.65 \pm 0.18$
                                                (0.73)~~
                                                                                                       &~~$0.05 \pm  0.01$ (0.05)~~ \\
  {+CMB+DES-Dovekie} & 3.1 & 2.8 & $0.62 \pm 0.18$ (0.69) & $0.05 \pm
                                                           0.01$
                                                           (0.05) \\
   \hline
\end{tabular}
\end{table}

This simple model is integrated into the hybrid {\tt CLASS}~\cite{Blas:2011rf,Lesgourgues:2011rh} and {\tt
  COBAYA}~\cite{Torrado:2020dgo} MCMC pipeline to identify the parameter space regions empirically supported
by DESI + CMB + SN Ia data. The results taken
from~\cite{Bedroya:2025fwh} are encapsulated in
Tables~\ref{tabla5} and \ref{tabla6}, and can be summarized as
follows. Actually, the statistical analysis compares two scalar field
models, distinguished by the sign of $c_1$ (positive or negative) with
$c_2 \geq 0$. A positive (negative) value of $c_1$ corresponds to a model where the
fading of dark sector (FDS) accompanies a decreasing (increasing) tower scale, which in the dark dimension scenario reflects an expanding (contracting) fifth dimension.

In both models, the scalar field moves toward larger values during the
dark matter-dominated era, causing dark matter mass to
decrease. However, the models depart from their shared trajectory
during the transition to dark energy domination: if $c_1 > 0$, the
dark matter mass continues to drop at a different rate, whereas if
$c_1 < 0$, the scalar field reverses, causing dark matter mass to
increase at late times. Tables~\ref{tabla5} and \ref{tabla6} show
that these models perform similarly to the CPL parametrization
compared to $\Lambda$CDM, with a slight preference for the $c_1 < 0$
case. Notably, all dataset combinations including CMB and DESI
consistently favor a non-zero $c_2 = 0.05 \pm 0.01$. Moreover,
the best-fit of $c_{2}$ is remarkably consistent with the $c_2 <0.2$
upper limit~\cite{Kesden:2006zb} imposed by the lack of fifth-force
detection in the dark sector. The last three rows in
Tables~\ref{tabla5} and \ref{tabla6} stand for the recalibrated SN
Ia samples, with the CPL
parametrization favored at more than $3\sigma$ and a comparable
significance is shown by the FDS model.
While the significance for some of the recalibrated SN Ia datasets
decreases, the overall message remains consistent and robust, showing
greater agreement between samples. Furthermore, the best-fit parameters $c_{1}$ and $c_{2}$ remain consistent across all recalibrated SN Ia samples~\cite{Bedroya:2025fwh}.

To connect with the phenomenological discussion in Sec.~\ref{sec:5a}, we begin by writing the CDM density as
$\rho_{\rm CDM} = n \, m(\phi)$. Its time
derivative is
\begin{equation}
  \dot{\rho }_{\rm CDM}=\dot{n} \, m(\phi )+n \, \dot{m}(\phi) \, .
\label{dotrho}  
\end{equation}
Given the mass function $m(\phi) = m_0 \, e^{-c_2\phi}$, its time derivative is
\begin{equation}
\dot{m}(\phi )=\frac{dm}{d\phi }\dot{\phi
}= -c_2 \, m_{0} \, e^{-c_2\phi }\, \dot{\phi }=-c_2 \, \dot{\phi } \, m(\phi) \, .
\label{dotm}
\end{equation}
Assuming that the total particle number is
conserved,\footnote{Using the decay rates of the KK gravitons given
  in~\cite{Obied:2023clp} it is easy to check that this is indeed a
good approximation.} the number density satisfies the standard continuity
equation \(\dot{n} + 3Hn = 0\), which means
\begin{equation}
\dot{n}=-3Hn \, .
\label{dotn}
\end{equation}
Substituting (\ref{dotm}) and (\ref{dotn})
 back into the (\ref{dotrho}) yields
\begin{equation}
\dot{\rho }_{\rm CDM}= -3H\, n \, m(\phi ) -n \, c_2\, \dot{\phi
  } \, m(\phi) =-3H \, \rho_{\rm CDM}-c_2 \, \rho _{\rm CDM} \, \dot{\phi } \, .
\end{equation}
Rearranging these terms provides the modified continuity
equation
\begin{equation}
\dot{\rho }_{\rm CDM}+3H\rho _{\rm CDM}=-c_2\rho
_{\rm CDM}\dot{\phi } \, .
\label{dotrhoc2}
\end{equation}

Direct comparison of (\ref{dotrhoQ}) and (\ref{dotrhoc2}) yields $Q = c_2\rho
_{\rm CDM}\dot{\phi}$, which implies $\beta \equiv
c_2$. Additionally, Tables~\ref{tabla4}, \ref{tabla5}, and \ref{tabla6} demonstrate that for standard
exponential scalar field mass dependence, MCMC likelihood fits to the
combined DESI + CMB + SN Ia datasets are practically independent of the potential's shape, $V(\phi)$.

We now turn to evaluate the FDS model against swampland conjectures. By
setting $d=4$ in  (\ref{TCCasym}) and (\ref{asymmlightest}), we
establish the asymptotic lower limits for $c_1$ and $c_2$ 
\begin{equation}
  (c_1,c_2)_{\infty} \geq \left(\sqrt{2}, \sqrt{2}/2\right) \, .
\label{c1c2asym}
\end{equation}  
The best-fit exponents given in Tables~\ref{tabla5} and \ref{tabla6}
do not satisfy the lower limit (\ref{c1c2asym}) for asymptotic field space behavior, but are of the same order of magnitude. In contrast, $c_{1}$ satisfies
the TCC bound (\ref{TCCbulk2}) required for the interior of the moduli
space, i.e. $c_1 \agt \gamma' \simeq 0.8$.

Because the MCMC likelihood fits to the combined DESI, CMB, and SN Ia
data show little sensitivity to the potential's shape for CQ, it is
worth exploring whether a potential that satisfies the TCC in the
asymptotic regime can also produce the appropriate slope away from
this limit, within the field space interior. We base our exploration
on the + DES-Dovekie fit shown in Table~\ref{tabla5}. Along this line, the
$S$-dual potential with $\varkappa = \sqrt{2}$
satisfies
\begin{equation}
  \nabla \log V(\phi )=-0.81\pm 0.19 \, ,
\end{equation}
inside the interval $\phi \in [-0.70,-0.33]$. Furthermore, throughout this field range, the integrated square error between the
$S$-dual potential and the exponential form stays below 2.5\%, see Appendix. This concomitance suggests that (\ref{Vsdual}) with asymptotic slope $\varkappa = \sqrt{2}$ (to satisfy the TCC bound as the moduli space distance approaches infinity) provides a valid global description of the potential that
locally behaves as an exponential.

\section{Key Takeaways and the Path Ahead}
\label{sec:6}

The synergy between quantum gravity and experimental cosmology is at
an all-time high, fueled by observational data validating predictions
from swampland
conjectures. Particularly, the 2024-2025 results from DESI-BAO measurements have
provided evidence suggesting that combined data from CMB, SN Ia, and
BAO now favor dynamical dark energy over a cosmological constant
$\Lambda$. With a confidence level over \(3\sigma\) and refined SN calibrations, a consistent picture is emerging across all major datasets. DESI results have
prompted intense theoretical activity, challenging the long-dominant
$\Lambda$CDM model and sparking a surge in beyond SM physics
proposals. Notwithstanding the expansive literature regarding DESI 
findings, this review is restricted to string-inspired scenarios,
examining the constraints mandated by the Swampland program. The
following is a representative, rather than exhaustive, list of
alternative data interpretations outside our current
scope~\cite{Jiang:2024xnu,Figueruelo:2026eis,Yang:2025uyv,Santos:2025wiv,Wang:2026sqy,Sahlu:2026bsa,Paliathanasis:2026ymi,Paliathanasis:2025mvy,Khodadi:2025gsm,Li:2025vqt,Dasgupta:2025ypg,Luciano:2025fqg,Luciano:2025dhb,Luciano:2025ykr,Gialamas:2025pwv,Paliathanasis:2025xxm,An:2025vfz,Cai:2025mas,Luciano:2025elo,vanderWesthuizen:2025iam,Scherer:2025esj,Luciano:2025hjn,Li:2025cxn,Yang:2025mws,Teixeira:2025czm,SPT-3G:2025vyw,Li:2025ops,Mazumdar:2025smh,Brax:2025ahm,Paliathanasis:2025kmg,Wu:2025vfs,Shen:2025cjm,Brassel:2025cub,Li:2026hwq,Chaudhary:2025vzy,Maki:2026zrn,Montefalcone:2026iga,Najafi:2026kxs,deCruzPerez:2026mkg,Woo:2026ice,Jing:2026ymp,Addazi:2026ulq,Kibris:2026cqq}.

On the theoretical front, the Swampland program has introduced a new
perspective on naturalness. This framework outlines several key phenomenological features of UV/IR mixing, including: {\it (i)}~the naturalness of a small vacuum energy 
  characterized by an exponentially suppressed potential $V/M_p^4 \ll 1$, {\it (ii)}~the tendency for the slope of
the potential to be proportional to its value, $|\nabla
V| \sim V$; and {\it (iii)}~the emergence of a tower of weakly coupled states with
exponentially light masses in
  regions where the potential is small, $m_{\rm tower}/M_p \ll
  1$. These constraints restrict the scale of inflation to $\eta <
  10^9~{\rm GeV}$ and limit the tensor-to-scalar ratio, thereby
  pushing associated primordial gravitational waves out of the range
  of experimental detection.
  Essentially, swampland conjectures
  introduce fundamental tensions that challenge the validity of the
  inflationary framework. On the plus side, compelling alternatives exist,
  especially within string theory frameworks. In addition, swampland
  constraints have shaped the 
  landscape of quintessence potentials, favoring those compatible with
  DESI data. These data also provided initial indications of non-gravitational interactions within the dark sector.

Deciphering the dark sector will require a unified strategy bridging observation, data-analysis, and theory. Crucially, while initial DESI findings are intriguing, a prudent approach is
necessary.  Further validation from the DESI Y5 data and concurrent
surveys (such as {\it Euclid}~\cite{Euclid:2025bxg},
Rubin/LSST~\cite{LSST:2008ijt}, and the Nancy
Grace Roman Telescope~\cite{Spergel:2015sza}) is essential to determine if the evidence against $\Lambda$CDM
will hold firm or gains even more traction, potentially ushering in a golden era of experimental string theory.

\acknowledgments{This review is dedicated to Ignatios Antoniadis
  for his 70th birthday, celebrating a lifetime of curiosity and
  contributions to theoretical physics. We are grateful for the
  insights gained from discussions with Ignatios Antoniadis, Alek
  Bedroya, Niccol\`o Cribiori, Jules Cunat, Eleonora Di Valentino, Muldrow Etheredge, Arda Hasar, Mustapha Ishak-Boushaki, Severin L\"ust, Joaquin Masias, Georges Obied, Karem Pe\~nal\'o Castillo, Marco Scalisi, Jorge Fernandez Soriano,
  Cumrun Vafa, and David Wu.  The work of L.A.A. is supported by the U.S. National Science
  Foundation (NSF Grant PHY-2412679), he 
 extends his appreciation to the Harvard Swampland Initiative for their warm hospitality and for providing a stimulating environment for productive discussions.
  The work of D.L. is supported by the German-Israel-Project (DIP) on
Holography and the Swampland.}

\section*{Appendix}

In this Appendix we adopt the continuous least-squares method to
analyze the similarities of the $S$-dual potential and the exponential
potential. We use as reference the MCMC likelihood fit from the
combined dataset DESI DR2 +
CMB + DES-Dovekie, in which $c_1 = -0.81 \pm 0.19$, see
Table~\ref{tabla5}. As the $S$-dual potential is invariant under reflection across the origin, we examine the positive branch.

\begin{figure}[tpb]
\postscript{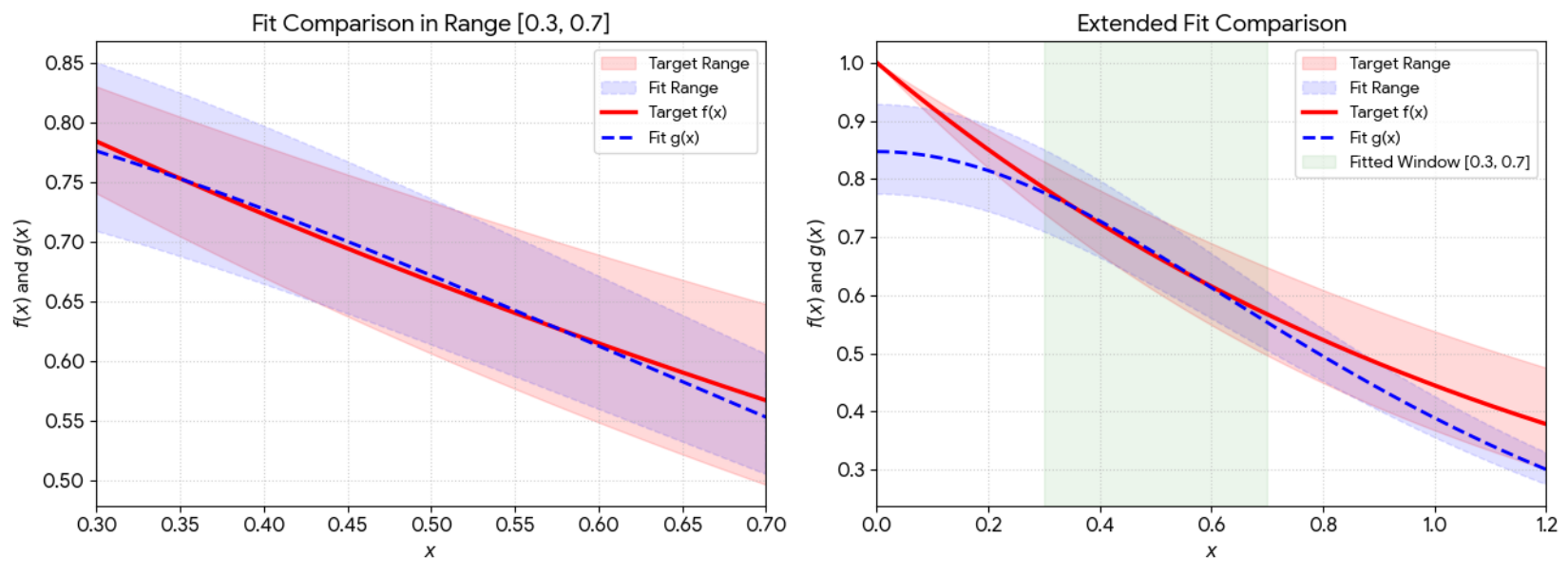}{0.95}
\caption{Comparison of the target exponential function $f(x)$ with the
  optimal least-squares fit $g(x)$ over the interval $x \in [0.3,
  0.7]$ (left) and in the extended region (right). \label{fig:4}}
\end{figure}

To minimize the integrated squared error,
\begin{equation}
E(A) = \int_{x_1}^{x_2} \left[f(x) - g(x)\right]^2
\, dx \,,
\end{equation}
between
\begin{equation}
f(x) = \exp(-c_1 x) \quad {\rm and} \quad g(x) = A \, {\rm
  sech}(\sqrt{2} \, x)
\end{equation}
in the interval $[x_1,x_2]$, we solve $dE/dA = 0$, yielding
\begin{equation}
A=\frac{\int _{x_1}^{x_2}\exp (-c_1x) \ {\rm sech}(\sqrt{2} \, x)
  \,dx}{\int _{x_1}^{x_2} {\rm sech}^{2}(\sqrt{2} \, x)\,dx} \, .
\label{Afit}
\end{equation}    
The optimal amplitude that minimizes the least-squares error over the interval $x \in [0.3, 0.7]$ is $A=0.8471$ for the
nominal case $c_1 = 0.81$. Under the parameter range $c_1 = 0.81 \pm
0.19$, the optimal values span from $A=0.7737$ to
$A=0.9279$.\footnote{The denominator of (\ref{Afit}) is 0.2523. For the nominal
  case, the numerator is 0.2137, for the lower bound 0.2341, and for
  the upper bound 0.1952.} The best fit is shown in Fig.~\ref{fig:4}.

\begin{figure}[tpb]
\postscript{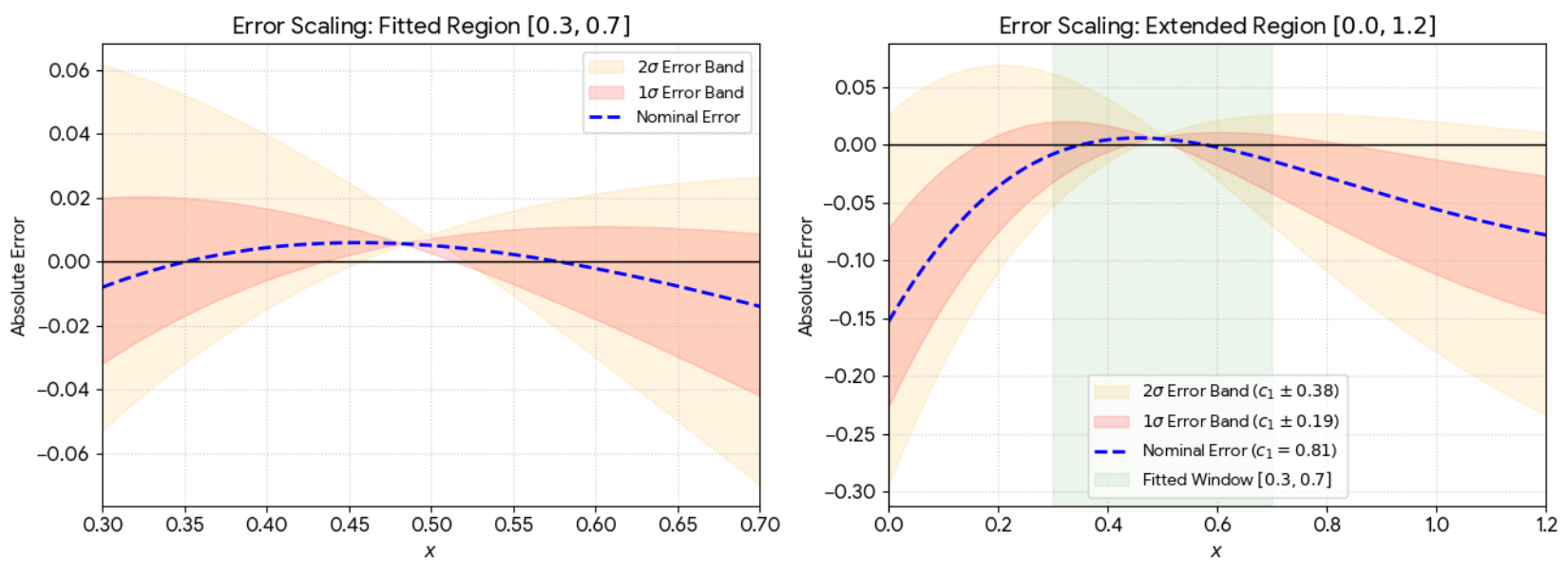}{0.95}
\caption{Absolute error over the interval $x \in [0.3,
  0.7]$ (left) and in the extended region (right).
\label{fig:5}}
\end{figure}

The absolute error shown in Fig.~\ref{fig:5} is computed by subtracting the target
function $f(x)$ from the optimized fit function $g(x)$ at every point
$x$. The red shaded error variation band tracks the absolute error
across the whole parameter range. At each individual point $x$, we
calculate the absolute error for all three cases $c_1 = 0.62$, $c_1 =
0.81$, and $c_1 = 1.00$. We then find the minimum and maximum error
values at that exact $x$ coordinate to determine the top and bottom
boundaries of the shading. The error stays exceptionally low throughout the window, remaining well under $2.5\%$ across the entirety of the optimization span.
For the extended region, the absolute error climbs sharply to over $-0.15$ because the boundary constraints were completely ignored during optimization.

For a continuous interval $[x_1, x_2]$, the continuous RMSE is defined
as the square root of the average squared error:
\begin{equation}
\text{RMSE}=\sqrt{\frac{1}{x_{2}-x_{1}}\int _{x_{1}}^{x_{2}}
  \left[g(x)- f(x)\right]^{2}\,dx} \, .
\end{equation}
Evaluating the integration for the optimized fits yields the following
performance values:
\begin{enumerate}[noitemsep,topsep=0pt]
\item Inside the fitted interval, i.e. $x \in [0.3, 0.7]$,
\begin{itemize}[noitemsep,topsep=0pt]
\item Nominal case $c_1 = 0.81$ $\to {\rm RMSE} \approx
  0.0016$.
\item Lower bound $c_1 = 0.62$ $\to {\rm RMSE} \approx
  0.0219$.
\item Upper bound $c_1 = 1.00$ $\to$ ${\rm RMSE} \approx 0.0163$.
\end{itemize}
\item Outside in the extended interval, i.e. $x \in [0.0, 1.2]$,
\begin{itemize}[noitemsep,topsep=0pt]
\item  Nominal case $c_1 = 0.81$ $\to {\rm RMSE} \approx 0.0435$.
\item Lower bound $c_1 = 0.62$ $\to {\rm RMSE} \approx 0.0521$.
\item Upper bound $c_1 = 1.00$ $\to {\rm RMSE} \approx 0.0684$.
\end{itemize}
\end{enumerate}
Note that the RMSE increases by over 25 times when moving from the fitted interval to the extended interval. This jump perfectly quantifies the visual ``error ballooning''  driven by the mismatch near $x=0$.

\end{document}